\newif\ifshort
\shortfalse

\documentclass[journal,a4paper]{IEEEtran}
\IEEEoverridecommandlockouts

\usepackage[utf8]{inputenc}
\usepackage[T1]{fontenc}
\usepackage{url}
\usepackage{bm}
\usepackage{ifthen}
\usepackage{cite}
\usepackage{graphicx}
\usepackage{todonotes}
\usepackage{soul}
\usepackage[cmex10]{amsmath} % Use the [cmex10] option to ensure complicance
\usepackage{breakurl}
\usepackage[breaklinks]{hyperref}

\usepackage{comment}
\usepackage{amsthm}
\usepackage{amsmath,amssymb,amsfonts}
\usepackage{graphicx}
\usepackage{subfig}
\usepackage{algpseudocode}
\usepackage{algorithm}
\usepackage{footmisc} % footref
\usepackage{enumitem} % Horizontal Spacing in List Items
\usepackage{multirow}
\usepackage{glossaries}
\setacronymstyle{long-short}
\glsdisablehyper
\newacronym{urllc}{URLLC}{Ultra-Reliable and Low-Latency Communications}
\newacronym{xurllc}{xURLLC}{neXt generation Ultra-Reliable and Low-Latency Communications}
\newacronym{sls}{SLS}{Sector Level Sweep}
\newacronym{llc}{LLC}{Low Latency Communications}
\newacronym{mmwave}{mmWave}{Millimeter-wave}
\newacronym{iab}{IAB}{Integrated and Access Backhaul}
\newacronym{los}{LOS}{Line-Of-Sight}
\newacronym{edca}{EDCA}{Enhanced Distributed Channel Access}
\newacronym{rlnc}{RLNC}{Random Linear Network Coding}
\newacronym{snr}{SNR}{Signal-to-Noise Ratio}
\newacronym{fec}{FEC}{Forward Error Correction}
\newacronym{ldpc}{LDPC}{Low-Density Parity-Check}
\newacronym{rcpi}{RCPI}{Received Channel Power Indicator}
\newacronym{per}{PER}{Packet Error Rate}
\newacronym{rtt}{RTT}{Round Trip Time}
\newacronym{bdp}{BDP}{Bandwidth-Delay Product}
\newacronym{ack}{ACK}{Acknowledgment}
\newacronym{nack}{NACK}{Negative Acknowledgment}
\newacronym{pbss}{PBSS}{Personal Basic Service Set}
\newacronym{sta}{STA}{Station}
\newacronym{twamp}{TWAMP}{Two-Way Active Measurement Protocol}
\newacronym{cp}{CP}{Channel Profile}
\newacronym{sp}{SP}{Single-Path}
\newacronym{mp}{MP}{Multi-Path}
\newacronym{pcp}{PCP}{Personal basic service set Control Point}
\newacronym{rssi}{RSSI}{Received Signal Strength Indicator}
\newacronym{ntp}{NTP}{Network Time Protocol}

\usepackage{mathtools}

\usepackage{xcolor}
\definecolor{DarkGreen}{rgb}{0.1,0.5,0.1}
\definecolor{DarkRed}{rgb}{0.5,0.1,0.1}
\definecolor{DarkBlue}{rgb}{0.1,0.1,0.5}
\definecolor{DarkPurple}{rgb}{0.5,0.2,0.5}
\definecolor{DarkTurquoise}{rgb}{0.1,0.5,0.5}

\definecolor{beaublue}{rgb}{0.74, 0.83, 0.9}
\definecolor{coolblack}{rgb}{0.0, 0.18, 0.39}
\definecolor{apricot}{rgb}{0.98, 0.81, 0.69}
\definecolor{burntorange}{rgb}{0.8, 0.33, 0.0}
\definecolor{blue-violet}{rgb}{0.54, 0.17, 0.89}
\definecolor{byzantium}{rgb}{0.44, 0.16, 0.39}
\definecolor{brilliantrose}{rgb}{1.0, 0.33, 0.64}
\definecolor{cerisepink}{rgb}{0.93, 0.23, 0.51}
\definecolor{cobalt}{rgb}{0.0, 0.28, 0.67}
\definecolor{bostonuniversityred}{rgb}{0.8, 0.0, 0.0}

\usepackage{graphicx}
\usepackage{booktabs} % To thicken table lines
\newif\ifcomen
\comentrue
\newcommand{\off}[1]{}

\begin{document}

\title{Ultra-Reliable Low-Latency Millimeter-Wave Communications with Sliding Window Network Coding \thanks{E. Dias, D. Raposo and T. Ferreira are with Instituto de Telecomunica\c{c}\~oes, Aveiro, Portugal (e-mail:{eurico.omdias, dmgraposo, tania.s.ferreira}@av.it.pt). H.~Esfanaizadeh and M.~Médard are with the EECS Department, Massachusetts Institute of Technology (MIT), Cambridge, MA 02139 USA (email: {medard, homaesf}@mit.edu). A.~Cohen is with the Faculty of ECE, Technion, Israel (e-mail: alecohen@technion.ac.il). M. Lu\'is is with Instituto Superior de Engenharia de Lisboa and Instituto de Telecomunica\c{c}\~oes, Portugal (e-mail:nmal@av.it.pt). S. Sargento is with the University of Aveiro and Instituto de Telecomunica\c{c}\~oes, Portugal (e-mail:susana@ua.pt).}}

\author{
    \IEEEauthorblockN{Eurico Dias, Duarte Raposo, Homa Esfahanizadeh, Alejandro Cohen,\\ Tânia Ferreira, Miguel Luís, Susana Sargento, and Muriel M\'edard}\\
}
\maketitle

\begin{abstract}
Ultra-reliability and low-latency are pivotal requirements of the new 6th generation of communication systems (xURLLC). Over the past years, to increase throughput, adaptive active antennas were introduced in advanced wireless communications, specifically in the domain of millimeter-wave (mmWave). Consequently, new lower-layer techniques were proposed to cope with practical challenges of high dimensional and electronically-steerable beams. The transition from omni-directional to highly directional antennas presents a new type of wireless systems that deliver high bandwidth, but that are susceptible to high losses and high latency variation. Classical approaches cannot close the rising gap between high throughput and low delay in those advanced systems. In this work, we incorporate effective sliding window network coding solutions in mmWave communications. %, which can track the current pattern of losses in the high-frequency channels and significantly improve the performance.
While legacy systems such as rateless codes improve delay, cross-layer results show that they do not provide low latency communications (LLC - below 10 ms), due to the lossy behaviour of mmWave channel and the lower-layers' retransmission mechanisms. On the other hand, fixed sliding window random linear network coding (RLNC) is able to achieve LLC, and even better, adaptive sliding window RLNC obtains ultra-reliable LLC (\gls{urllc} - LLC with maximum delay below 10 ms with more than 99\% success rate).

\end{abstract}

%%%%%%%%%%%%%%%%%%%%%%%%%%%%%%%%%%%%%%%%%%%%%%%%%%%%%
\section{Introduction}
%%%%%%%%%%%%%%%%%%%%%%%%%%%%%%%%%%%%%%%%%%%%%%%%%%%%
\gls{mmwave} networks enable multi-gigabit-per-second data rates between $57$~GHz and $64$~GHz, the so-called V-Band, that uses the unlicensed spectrum available worldwide. It is an attractive option for \gls{iab}, which is part for the new generation of communications - 6G- to reduce deployment expenses of fiber optics with the increase of connection density \cite{9397776}. However, these frequency bands have been heretofore mostly idle because mmWave communications suffer from strong path loss, and heavy propagation challenges with obstacles, rain and atmospheric absorption, making them only suitable for short and \gls{los} communications. Recent advances in the use of small antenna arrays, capable of forming highly-dimensional and electronically-steerable beams, and beamforming techniques like the \gls{sls} \cite{nitsche2014ieee}, can partially ameliorate the effects of propagation characteristics \cite{Wang2018}, but with associated complexity and costs.

The \gls{llc} and \gls{urllc} are two integral parts of modern communications systems~\cite{9826826}. In-order delivery delay is the main target for both \gls{llc} and \gls{urllc}, and in addition, \gls{urllc} would require the bulk of the packets to be delivered in a timely fashion to their destinations, \emph{i.e.}, with a failure probability of less than \begin{math}1-10^{-5}\end{math}, and within a latency of 1~ms for 32~bytes and 3-10~ms for 300~bytes \cite{3gpp.38.913}.

% transmission with the targeted in order delay deadline. \textcolor{red}{[more info here]}

%Such change results in the use of beamforming techniques, that introduce a high overhead procedure in the search space, which scales based on the product of the sender-receiver beam resolution.  However, a hybrid backhaul that consists of fiber links and mmWave links is still challenging,
The challenges of mmWave are particularly salient when we seek to use them, as would be the case in IAB, for the \gls{xurllc} in 6G services (e.g., tactile internet, VR/AR and intelligent transportation). The lossy nature of mmWave introduces new challenges in MAC and transport layers, such as link quality assessment, rate adaptation and bufferbloat~\cite{Ren2021}. Error-control mechanisms at the transport layer can tackle the dramatic path loss, but could also lead to blockage-induced timeouts, introducing unnecessary data retransmission (e.g., TCP). The use of \gls{edca} with the in-order block acknowledgment scheme can prevent extra unnecessary retransmissions \cite{Dahhani2019}, but at cost of lowering the MAC goodput (due to holding off the link layer sliding window).  The mismatch between current mmWave mechanisms - MAC and transport layers - leads to higher overall delays, that can not be presented in \gls{urllc} scenarios.  %S error-correction codes offer a high level of error-tolerance at a cost of inducing a notable latency \cite{LubMitShoSpiSte1997,huffman2021concise}, which intuitively is in contradiction with the philosophy of using mmWave for a faster communication.

Several techniques have been used to correct failures in the wireless channels, e.g., rateless erasure codes  \cite{shokrollahi2006raptor,luby2002lt}, which were recently deployed by Verizon \cite{Verizon}, systematic codes \cite{cloud2015coded}, and streaming codes \cite{joshi2012playback,joshi2014effect}. %None of those solutions is tracking the varying channel conditions to provide a suitable trade-off between throughput and delay of the communication.
In order to manage delay, transport protocols commonly use windowing schemes, such as TCP \cite{cerf1974protocol,cerf1974specification}. Combining windowing with coding can be done with Random Linear Network Coding (SW-RLNC) either in a fixed way \cite{ho2006random,LunMedKoeEff2008} (F-SW-RLNC) or in an adaptive way \cite{cloud2015coded,yang2014deadline} (A-SW-RLNC). Recently, a causal A-SW-RLNC scheme was proposed  in \cite{cohen2020adaptive,cohen2020adaptiveMPMH,michel2022flec}. The main idea is to track the channel state to adjust the size of the window of packets used to form the RLNC-coded packet in a causal fashion. This feature adaptively tunes the redundancy ratio and error correction capability of the coding solution to obtain the desired delay-throughput trade-off.

\begin{figure}[t]
        \includegraphics[width=0.5\textwidth]{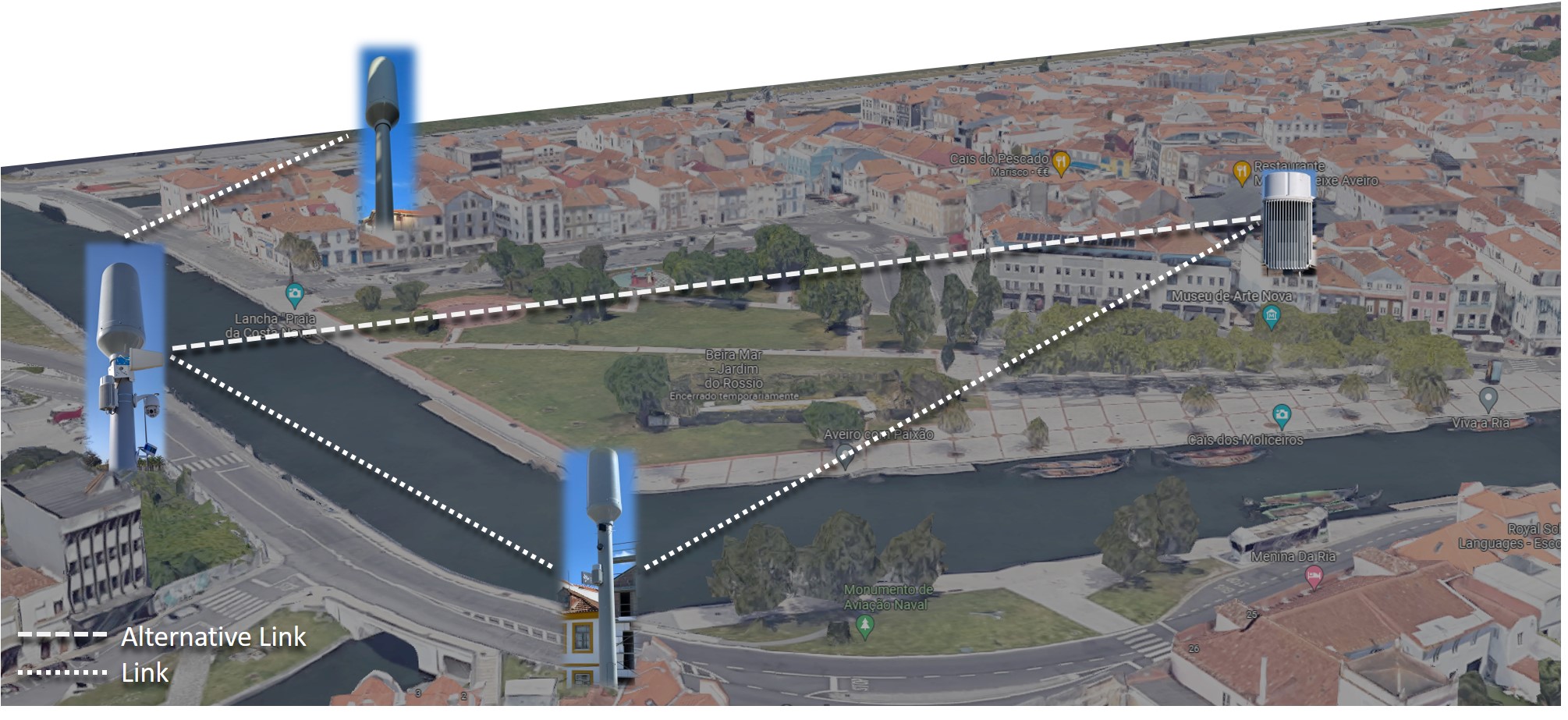}
        \caption{Aveiro Living Lab mmWave infrastructure (Portugal): Four mmWave antennas are connected with \gls{los} links \cite{Aveiro_SmartCity}.}
        \label{fig:SmartCity}
\end{figure}

The contributions of this paper are as follows: We look into the unique dynamic behaviour of mmWave communication environment, and propose how to use SW-RLNC, both F-SW-RLNC and A-SW-RLNC versions of \cite{cohen2020adaptive,cohen2020adaptiveMPMH}, to capture rapid changes and mitigate the high losses that are intrinsic to mmWave, for \gls{llc} and \gls{urllc}. In particular, we study the interplay of cross-layer solutions, and propose how one can mitigate the large delay caused by the lower-layer error control mechanisms. We further, theoretically, study the SW-RLNC scheme in terms of its maximum delay guarantees over the transport-layer mmWave channel model. This worst-case analysis plays an important role in \gls{urllc} compared to the customary average analyses that are performed in the recent literature. We evaluate the performance of our solution over our mmWave testbed deployed in the city of Aveiro (Fig.~\ref{fig:SmartCity}). For our study, we have collected a representative dataset in the mmWave backhaul network, and the obtained results show that A-SW-RLNC scheme can obtain \gls{urllc}, with a maximum delay below 10 ms with more than 99\% success rate.

The remaining of this paper is organized as follows: Section~\ref{sec:overview} describes the mmWave technology - PHY and MAC layer mechanisms -, the system model and problem formulation. Section~\ref{sec:coding} presents different RLNC approaches, and in specific, the  A-SW-RLNC. Section~\ref{sec:experiments} presents the experimental network test scenario, the methodology, and the experimental results by comparing various RLNC approaches in terms of their key performance metrics, \emph{i.e.}, throughput, mean in-order delay, and maximum in-order delay. Section~\ref{sec:theoretical} is dedicated to the theoretical validations of the results: mathematical modeling of the transport-layer mmWave channel and bounding the in-order delivery delay. Finally, Section~\ref{sec:conclusions} summarizes the findings and discusses future research directions.

%\ale{{\bf Gap}: Since the high loss pattern in mmWave, error correction codes in the physical layer, are used with high redundancy to get reliability. However, we pay using those solutions in advance with a high loss of rate.\\
%{\bf Main contribution}: Using network coding solutions for error correction (in the higher layers?), we can significantly reduce the loss in rate. We will show, incorporating different network coding algorithms in multi-path mmWave channel, that as the network coding solution can better track the current situation of the frequency channels, we can significantly improve the performance.}

%\dr{{\bf Gap}: Another important fact is that the small changes in beam steering may cause a lot of losses in transport protocols like TCP. Present here the commitment that we pay in the MCS and the FEC used in the lower layers}

%\newpage

%%%%%%%%%%%%%%%%%%%%%%%%%%%%%%%%%%%%%%%%%%%%%%%%%%%%%
\section{Problem Statement and Preliminaries}\label{sec:overview}
%%%%%%%%%%%%%%%%%%%%%%%%%%%%%%%%%%%%%%%%%%%%%%%%%%%%%
This section provides a technical background of mmWave technology, the system model and the problem formulation.
\subsection{MmWave Communications and Challenges}

The wireless propagation channel can vary significantly over time, greatly affecting the radio’s link quality. This is especially significant for technologies that operate in the mmWave band, such as IEEE 802.11ad \cite{802_11ad_standard}, as the higher frequencies cause higher susceptibility to the blockage.

To mitigate the negative impact of obstructions, Wigig-based COTS devices, such as the CCS Metnet nodes \cite{metnet_60g_datasheet}, employ an automatic mode to dynamically select the parameters of modulation and error correction based on the instantaneous \gls{snr} and error rate \cite{ccs_specs}. Specifically, devices can switch between four modes of operation, each having specific modulation and \gls{fec} schemes, and all utilize the 60~GHz carriers: a) control PHY (Modulation and Coding Scheme - MCS 0); b) Single carrier (MCS 1-12) PHY; c) OFDM (MCS 13-24) PHY; and d) low-power SC (MCS 25-31) PHY. This set of choices makes it possible to meet different performance requirements, depending on the usage scenario (e.g., low-interference, low-complexity, low-energy-consumption, etc.).

The 802.11ad PHY standard utilizes \gls{ldpc} codes with four different rates (1/2, 5/8, 3/4, and 13/16), with a fixed codeword length of 672 bits~\cite{schultz2013802}. Each modulation type combined with a specific code rate form a MCS. More details regarding the modulation schemes, code rates and data rates that are supported by each MCS are specified in Table \ref{tab:CCS_specs}. For the rate identification, the manufacturers can adopt a maximum frame size of 2000 bytes, and the calculations are performed in the second layer.

\begin{table}[b]
\centering
\caption{Different Modelation and Coding for Single carrier PHY mode supported by the CCS Metnet node \cite{802_11ad_standard}.}
\label{tab:CCS_specs}
\begin{tabular}{cccccc}
\toprule
\begin{tabular}[c]{@{}c@{}}\textbf{MCS}\end{tabular} & \textbf{MOD} & \textbf{Code rate} & \begin{tabular}[c]{@{}c@{}}\textbf{Min. RCPI}\\\textbf{(dBm)}\end{tabular} & \begin{tabular}[c]{@{}c@{}}\textbf{Min.}\\\textbf{SNR (dB)}\end{tabular} & \begin{tabular}[c]{@{}c@{}}\textbf{Layer 2}\\\textbf{Line} \\\textbf{(Mbps)}\end{tabular}  \\
\hline
0                                                      & DSSS                & 12                & -84.52                                                                          & -11                                                                            & 22.4                                                                                  \\
1                                                      & BPSK                & 1/2               & -73.72                                                                          & -0.2                                                                           & 308                                                                                   \\
2                                                      & BPSK                & 1/2               & -72.52                                                                          & 1                                                                              & 616                                                                                   \\
3                                                      & BPSK                & 5/8               & -71.32                                                                          & 2.2                                                                            & 770.4                                                                                 \\
4                                                      & BPSK                & 3/4               & -69.92                                                                          & 3.6                                                                            & 924                                                                                   \\
5                                                      & BPSK                & 13/16             & -69.02                                                                          & 4.5                                                                            & 1000.8                                                                                \\
6                                                      & QPSK                & 1/2               & -69.72                                                                          & 3.8                                                                            & 1232                                                                                  \\
7                                                      & QPSK                & 5/8               & -68.22                                                                          & 5.3                                                                            & 1540                                                                                  \\
8                                                      & QPSK                & 3/4               & -66.72                                                                          & 6.8                                                                            & 1848                                                                                  \\
9                                                      & QPSK                & 13/16             & -65.72                                                                          & 7.9                                                                            & 2002.4                                                                                \\
\bottomrule
\end{tabular}
\end{table}

\begin{figure}[t]
        \includegraphics[width=0.50\textwidth]{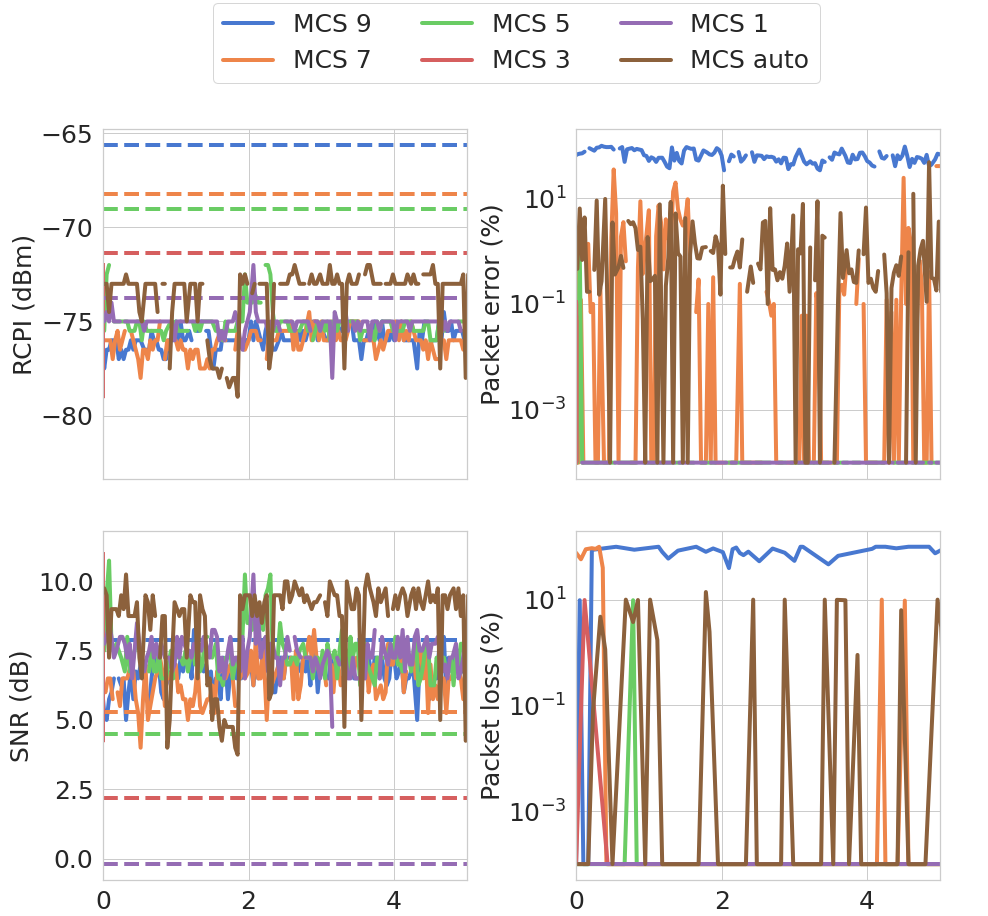}
        \caption{Measurements for PHY and MAC layer under blockage, for a time interval of 5 minutes. }
        \label{fig:noRLNC_obstructed_metrics}
\end{figure}

Fig.~\ref{fig:noRLNC_obstructed_metrics} presents a set of measurements performed in a mmWave outdoor setup, further described in Section~\ref{section:testbed}, considering an obstructed scenario. Several PHY layer metrics were recorded, such as, \gls{rcpi}, \gls{snr}, and \gls{per}, and one transport layer metric, packet loss rate, for different MCS modes (fixed and automatic). In addition, the figure also shows the minimum required levels of \gls{snr} and \gls{rcpi} reported by the manufacturer for maintaining each MCS (the dashed horizontal lines). As shown, under obstruction, the \gls{rcpi} signal level drops to a lower value than the defined threshold for almost all fixed MCSs (with the exception of MCS 1). However, the SNR requirement is still being fulfilled most of the time. Note that sudden decreases of the \gls{rcpi} and \gls{snr} occurred due to slight obstacle movements. Moreover, the figures show that using higher modulation modes in an obstructed scenario may lead to packet error rates up to 90\% in the PHY layer, and packet loss rate up to 100\% in the transport layer. On the other hand, automatic modulation adjustment leads to lower packet losses and packet error rates, under a static obstruction.

Still, switching MCS to a more robust scheme (\emph{i.e.}, with a lower code rate) whenever the signal quality drops may not be enough to ensure a reliable connection. This is because a sudden obstruction of the \gls{los} path can cause a significant decrease in the maximum achievable throughput, increasing the delay. This leads to several issues in the upper layers, such as link quality judgment, rate adaptation, and bufferbloat~\cite{Ren2021,Dahhani2019}. Link fluctuations will affect TCP performance when mmWave links switch between \gls{los} and non-\gls{los} links states, resulting in TCP retransmissions, an increase of the \gls{rtt}, and consequently, a decrease of the congestion window. The variability of the links also prevent some protocols to achieve the ultra-high bandwidth capacity of the mmWave links. The \gls{bdp} used to compute the optimal buffer size is difficult to estimate, resulting in high latency and jitter when using larger buffers to prevent packet losses.

Thus, algorithms that are faster to adapt to the conditions of the link have major advantages for this setting (e.g., A-SW-RLNC), which is the main scope of the current paper. We use an effective network coding solution that can be introduced in the transport layer to mitigate the losses and handle the high variations of delay caused by obstructions. This, in turn, lightens the requirements of the \gls{fec} mechanisms implemented at the PHY layer. This is especially useful in mmWave networks where blockage increases the delay, packet losses, and the number of TCP retransmissions.

%%%%%%%%%%%%%%%%%%%%%%%%%%%%%%%%%%%%%%%%%%%%%%%%%%%%%%%
\subsection{System Model and Problem Formulation}\label{subsec:system}

We consider a real-time slotted mmWave communication with feedback. In particular, a \gls{sp} communication setting is considered between two points, sender and receiver, and we assume that the data that needs to be transmitted consists of $N$ packets of the same size, \emph{i.e.}, $\{P_1,\dots,P_N\}$. At the $i$-th time step, the sender transmits a coded packet $E_i$ over the noisy mmWave forward channel. The receiver may acknowledge the sender by sending an \gls{ack} for any delivered coded packet over the noisy feedback channel. The delay between the first time data is transmitted and the time that the corresponding feedback is received is called \gls{rtt}. The transmission delay of coded packets $t_d$ is the time it takes for the sender to transmit one packet (push the packet into the medium), and the propagation delay $t_{\text{prop}}$ is the amount of time it takes for one packet to be received from the sender to the receiver and vice versa. We assume that the size of the feedback acknowledgment is negligible, and that the propagation delay can vary for any transmitted coded packet according to the channel's condition. Hence, the \gls{rtt} for each coded packet is equal to $t_d + 2t_{\text{prop}}({E_i})$, where $t_{\text{prop}}({E_i})\leq t_{\text{prop}}$. Let the timeout $t_{\text{o}}\geq 2t_{\text{prop}}$ to denote an adaptive parameter the sender may choose to declare the packets that were not delivered at the receiver. That is, for any coded packet transmitted, if an \gls{ack} is not received at the sender after $t_d + t_{\text{o}}$ time slots, the sender declares a \gls{nack} for the corresponding packet.

Our main performance metrics are defined as follows:

\noindent (1) {\bf Throughput $\bm\eta$}. This is defined as the rate, in units of bits per time slot, at which the information is delivered at the receiver. In this paper, we focus on a normalized throughput, denoted by $\eta$, which corresponds to the total number of information bits delivered to the receiver divided by the total amount of bits transmitted by the sender.

\noindent (2) {\bf In-order delivery delay of packets $\bm D$}. This is the difference between the time slot in which an information packet is first transmitted at the sender and the time slot in which the packet is decoded, in order, by the receiver.

Our goal in this setting is to maximize the throughput, while minimizing the in-order delivery delay of packets.

%\dr{I also discuss with Eurico if it's possible to use the metrics used before, when run directly in the mmWave testbed. Give me your feedback if it makes sense. Additional metrics that we can introduce: \paragraph{success-rate} for instance for a specific upload file size (e.g., 10.000 bytes), what will be the \% of delivery in each selected scenario: RLNC,  A-SW-RLNC, etc. \paragraph{download time} using the timeslot time (450$\mu$s), we can compute the download time. \paragraph{overhead} in terms of packets and processing time. In terms of packet overhead we can compare traditional transport protocols approaches (e.g., UDP), with the Network Coding algorithms. How much packets we need to transmit to send all the information in the channel.}

\section{Network Coding for Joint Scheduling and Coding}\label{sec:coding}

%[Place holder for traditional error correction mechanism and their shortcomings]

%\subsubsection{Error Correction via Network Coding}
%%%%%%%%%%%%%%%%%%%%%%%%%%%%%%%%%%%%%%%%%%%%%%%%%%%%%
This section elaborates on using RLNC as an error correction mechanism in the \textit{transport layer} between two points. This mechanism mitigates the rigid requirements of the physical layer error correction code to provide reliable communications for the worst mmWave channel condition. Thus, one can increase the performance in terms of throughput and delay as defined in Section~\ref{subsec:system}.

%\ale{We need to elaborate more about the main contribution for the described above in other sections.}

In classical RLNC schemes \cite{ho2006random,LunMedKoeEff2008}, each encoded packet $E_i$, where $i$ is a positive integer, that is transmitted over the lossy communication is a random linear combination of the original uncoded packets, i.e.,
\begin{equation}
    E_i=\sum_{j=1}^{N}\rho_{i,j}P_j,
\end{equation}
where the coefficients $\{\rho_{i,j}: i\in\{1,2,\dots\},j\in\{1,\dots,N\}\}$ are drawn from a sufficiently large field, and $N$ is the total number of original uncoded packets. In general, when the coefficients are randomly sampled from a large field, the receiver can decode the original packets once $N$ coded packets are received, for example using the Gaussian elimination technique.

Although classical RLNC schemes can achieve the desired communication rates in the realm of large $N$, it imposes a large latency to the system. This is because, to decode the first packet, at least $N$ coded packets need to be transmitted. Thus, some variations of RLNC have been studied in the literature to lower the latency \cite{luby2002lt,shokrollahi2006raptor,joshi2012playback,joshi2014effect,cloud2015coded,cohen2020adaptive}. In our proposed scheme, we compare part of these methods, which will be next described, over several mmWave settings in Section~\ref{sec:experiments}.

\subsection{Rateless RLNC (R-RLNC)}\label{subsec:brlnc}
In this variation, the sender's packets are split into non-overlapping blocks, called batches, each with size of $n$ packets. The batches are encoded and transmitted in order. For each batch, the encoded packets are random linear combinations of the packets within the same batch, and the ratio of the number of original packers $n$ and the number of encoded packets $m$ denotes the rate of the scheme. In a well-designed scheme, the receiver is able to recover the whole batch per receipt of its $n$ out of $m$ encoded packets. More precisely, let $E_{i}^k$ be the $k$-th encoded packet of the $i$-th batch, where $k\in\{1,\dots,m\}$, then
\begin{equation}
    E_{i}^k=\sum_{j=1}^{n}\rho_{i,j}^kP_{(i-1)n+j}.
\end{equation}
Here, it is assumed that the total number of packets is divisible by the block size. If not, one can easily use the zero-padding techniques. In this variation, the code designer in advance can try to manage the performance, in terms of throughput and latency trade-off, by choosing the size of $n$ and $m$.

When the sender does not receive an acknowledgement, by the end of transmitting the $n$-th coded packet, of a batch of at least $m$ packets, it starts sending another $n$ coded packets (with different coefficients) for the same batch. This process continues until the sender ensures that $m$ coded packets are received at the destination. Therefore, this RLNC scheme is by-definition a rateless code \cite{5978428}, and we call this scheme R-RLNC.

Recently, there are new solutions in the literature where the size of the $i$-th uncoded batch $n(i)$ and the size of the $i$-th coded batch $m(i)$ can be time-variant and adapted based on the channel estimation \cite{yang2014deadline,shi2015adaptive}. However, those solutions are only adaptive and reactive to the average packet loss probability. In mmWave communications, the channel conditions vary extremely fast; hence, although the above solutions are adaptive, one can pay in performance as those solutions do not track the specific erasure pattern of each packet and batch.

\subsection{Adaptive and Causal RLNC}\label{subsec:acrlnc}

This is an adaptive and causal variant (A-SW-RLNC method), as given in \cite{cohen2020adaptive}. In this method, at a time slot, according to the cumulative feedback information, the sender can decide either to transmit a new coded linear combination, i.e., \textit{new packet}, or repeat the last sent combination, \textit{same packet}. Here, \textit{same} and \textit{new} refer to the raw information packets contained in the linear combination, such that sending the same linear combination means that the raw information packets are the same but with different random coefficients. Thus, using a sliding window mechanism, the $i$-th coded packet can be described as follows,
\begin{equation}\label{eq:ac_coded}
    E_{i}=\sum_{j=w_{\min}}^{w_{\max}}\rho_{i,j}P_{j},
\end{equation}
where $w_{\min}$ corresponds to the oldest raw information packet that is not yet decoded, and $w_{\max}$ is incremented each time a new raw information packet is decided to be included in the linear combination by the sender.

The  A-SW-RLNC solution tracks the channel conditions, and adaptively adjusts the retransmission rates based on the channel quality and the feedback acknowledgments. For the channel estimation, the behavior of the channel parameters (i.e. erasure probability and its variance) is tracked using the feedback acknowledgements over time.  A-SW-RLNC envisions two different FEC mechanisms to add redundancy (retransmissions) and cope with the errors and failures, according to the channel status. The first one is $a priori$ and the second one is $a posteriori$, and they both interplay to obtain a desired throughput-delay trade-off. The first FEC mechanism is $a priori$, as it sends redundant packets in advance (before the failure occurs) according to the average estimation of the channel behavior. The second \gls{fec} mechanism is $a posteriori$, as it sends redundant packets according to the realization of errors, identified using the feedback information. It is through the second mechanism that the sender ensures that decoding is eventually possible at the receiver. We note that, the higher the number of $a priori$ \glspl{fec} is, the lower is the delay and the throughput, as it pro-actively recovers (possibly more than needed) for future lost coded packets. On the other hand, the higher the number of $a posteriori$ \glspl{fec} is, the higher is the delay and the throughput, as it only recovers for the needed lost coded packets at the cost of a delay proportional to \gls{rtt}. The way to adjust this trade-off is through an adaptive approach, which is described in detail in \cite{cohen2020adaptive}.

\section{Experimental Study and Performance Evaluation}\label{sec:experiments}
%%%%%%%%%%%%%%%%%%%%%%%%%%%%%%%%%%%%%%%%%%%%%%%%%%%%%
This section presents the experimental study of several transport layer communication solutions over mmWave links. We start by presenting the setup, characterizing the mmWave communication in terms of the packet loss and  \gls{rtt}, and then we discuss the performance of several communication solutions with respect to the \gls{llc} and \gls{urllc} requirements.

\subsection{Experimental Outdoor mmWave Setup}\label{section:testbed}

% TO DO: TÂNIA
\begin{figure}[t]
    \centering
    \includegraphics[width=1\columnwidth]{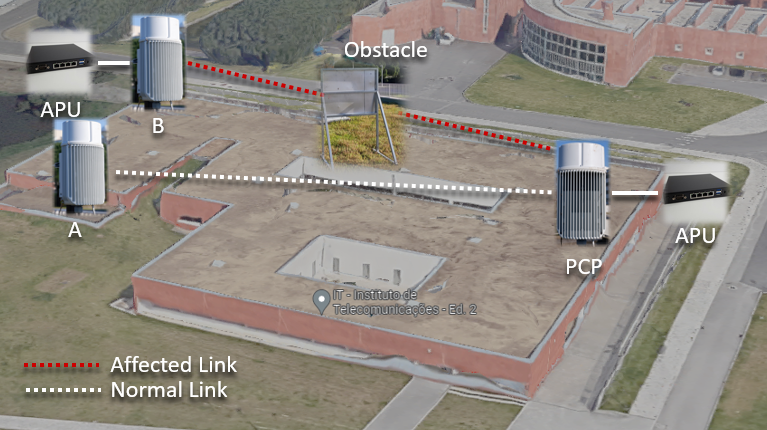}
    \caption{Experimental mmWave Network Test Scenario.}
    \label{fig:experimental_scenario}
\end{figure}

The mmWave network is composed of three CCS Metnet nodes \cite{metnet_60g_datasheet}, which are presented in Fig.~\ref{fig:experimental_scenario}. These nodes were deployed in an outdoor environment, specifically on the rooftop of Instituto de Telecomunica\c{c}\~oes, in Aveiro (Portugal), which allowed running tests under a fully controlled environment.
The deployed network adopts an architecture where the \gls{pcp} node has a wired connection to the core network. On the other hand, nodes A and B, the remote nodes, access the network through the radio links they establish with the node \gls{pcp}. For each node, there is a single board unit (APU) connected, that will communicate using the mmWave backhaul.

Furthermore, these nodes employ the standardized IEEE 802.11ad (WiGig) technology, which operates between the 57~GHz to 66~GHz unlicensed frequencies, to form a wireless 5G meshed backhaul capable of accommodating hundreds of gigabit traffic from the core network. Each device has four radio modules, each employing a 19~dBi beamforming steerable antenna that establishes directional links to cope with the increased attenuation at 60~GHz. For that purpose, the Wigig standard has proposed MAC and PHY layer enhancements which include support for directional communication through a process known as beamforming training, which allows determining the appropriate transmit and receive antenna sectors for communication between a given pair of stations. By employing beamforming techniques, the 300º horizontal field covered by each device is divided into 64 discrete sectors (with a 5º horizontal beamwidth) that can be used to concentrate the signal towards a specific direction.

\begin{figure}[t]
    \centering
    \includegraphics[width=0.7\columnwidth]{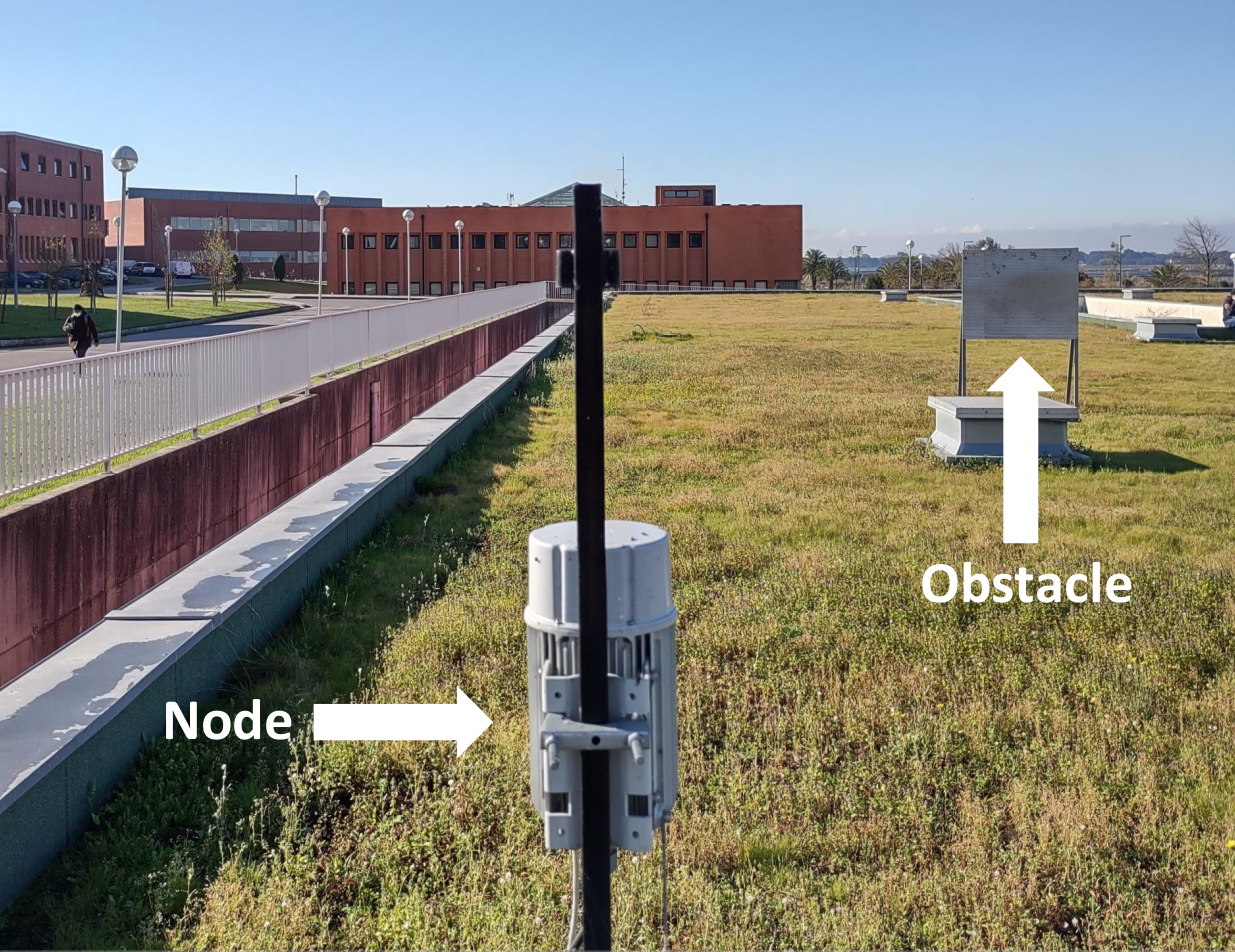}
    \caption{Blockage scenario with the metal obstacle.}
    \label{fig:obstacle}
\end{figure}

In the experimental scenario, a metallic obstacle was placed between two \glspl{sta} to simulate the blockage scenario represented in Fig.~\ref{fig:obstacle}. This obstacle was maintained in a fixed location for the entire test duration (25 minutes, i.e., 5 minutes per MCS mode). After the introduction of the obstruction, a significant reduction of the instantaneous \gls{rssi} and \gls{snr} was observed compared to the average values registered under non-obstructed operation (see Fig.~\ref{fig:noRLNC_obstructed_metrics}).

% TO DO: solve issue with the imports (package subcaption, subfigure command)!!
% \begin{figure}
%     \centering
%     \begin{subfigure}[]{.45\columnwidth}    %[t]{0.48\linewidth}        %% or \columnwidth
%         \centering
%         \includegraphics[width=\textwidth]{fig/network_node.jpg}
%         \caption{CCS Metnet node and its four radios.}
%         \label{fig:metnet_node_image}
%   \end{subfigure}%
%   \begin{subfigure}[b]{.45\columnwidth}   %[t]{0.48\linewidth}       %% or \columnwidth
%         \centering
%         \includegraphics[width=\textwidth]{fig/network_topology.png}
%         \caption{Testbed used throughout this work.}
%         \label{fig:network_topology}
%     \end{subfigure}%
%     \caption{Nodes and topology of the deployed outdoor testbed.}
%     \label{fig:network_and_node}
% \end{figure}

\subsection{Data Collection: Recording the mmWave Channel Profile}\label{test}

% \dr{{\bf Main topics}: This part could be for Eurico.
% How the datasets were generated regarding the losses and RTT. Also how the emulation part was built and what tools were used Kodo, OWAMP}

The dataset characterizing the mmWave channel behavior, \emph{i.e.}, the \gls{rtt} and the packet loss event at each time slot, was collected using the mmWave setup illustrated in Fig.~\ref{fig:experimental_scenario} and Fig.~\ref{fig:obstacle}. UDP traffic was generated, and at the same time, the \gls{rtt} and the state of the packet (being erased or not) were collected using the \gls{twamp} tool. The tool implements the standard defined in the RFC 5357\cite{rfc5357}, which is capable of performing two-way or round-trip measurements. Under the presented scenario, one of the Linux APU nodes was selected to be the sender, while the other was acting as a reflector. The \gls{twamp} tool was executed as a server on the reflector APU, up until the end of the dataset creation.

The unprocessed dataset consists of a 5-minute execution of the TWAMP tool for each MCS mode (\emph{i.e.}, MCS~3-MCS~6, and MCS~Auto).
%The limitation of the size and timeslot was determined by setting the interval between transmissions.
The time slot duration is 450~$\mu$s, which is obtained by trial and error as the maximum value without losing packets due to processing limitations. To minimize the measurement error, given that TWAMP one-way directional delays are clock sensitive, the system clocks on the APUs were synchronized using the \gls{ntp}, forced after each periodic measurement process.
%The raw and unprocessed dataset consists of a 5-minute execution of the TWAMP sender, for each MCS values from MCS3 through MCS6, fixed on the Metnet CCS radio nodes. The generation was also performed on the MCS Auto mode. The limitation of the size and time-slot was determined by sending a limited amount of packets, and by setting the interval between transmissions – the chosen value, obtained by trial and error, was the maximum the tool would support without losing packets due to its limitations on the APU, which was 450~$\mu$s \homa{This sentence is long a bit unclear. Can you please reword}.
In order to evaluate the channel behaviour and collect the metrics needed by the adaptive communication solutions, we retrieved from the unprocessed dataset a sequence of tuples, represented as (\gls{rtt}, packet loss) per time slot. We call this sequence a \textit{\gls{cp}}, which will be used in our proposed emulator described in the next subsection. In total, we obtained one channel profile for each MCS mode in $\{auto,3,4,5,6\}$.

%From a large set of metrics collected during the experiments we focused on the RTT and the packet loss, resulting in the \textit{Channel Profile (CP)} as follows
%\begin{equation}
%    CP_{MCS(m)}=\{sl_{m,1}, sl_{m,2}.., sl_{m,t}\},
%\end{equation}
%where $m=\{auto,3,4,5,6\}$ denotes the MCS mode, $t$ represents the time slot instant and $sl$ is the pair (RTT, packet loss) observed during the experiment.

\begin{figure*}[h]
	\centering
	\includegraphics[width=\textwidth,keepaspectratio]{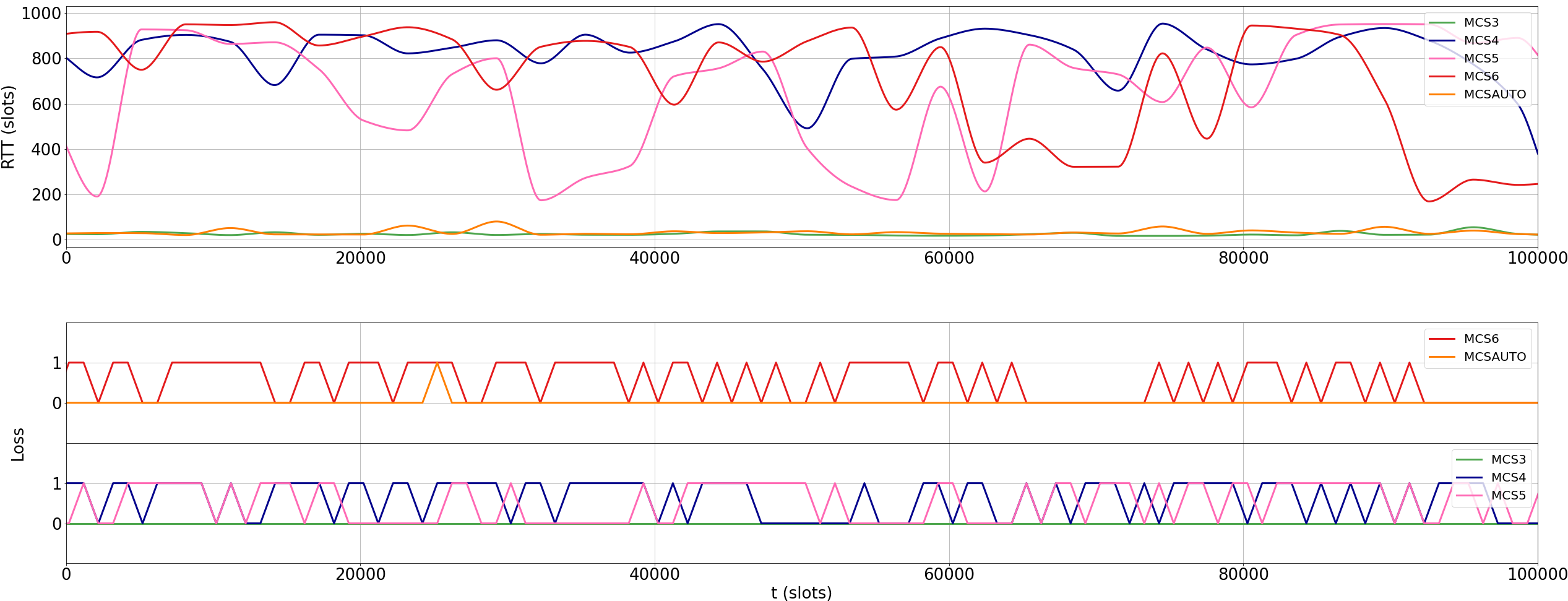}
	\caption{Channel profile - sequence of (RTT, packet loss) tuples, collected in the experimental mmWave network (in a blocked scenario with a metal obstacle), using different MCS modes, 3 to 6, and auto.}
	\label{fig:diff_curves}
\end{figure*}

\begin{figure*}[h]
	\centering
	\includegraphics[width=\textwidth,keepaspectratio]{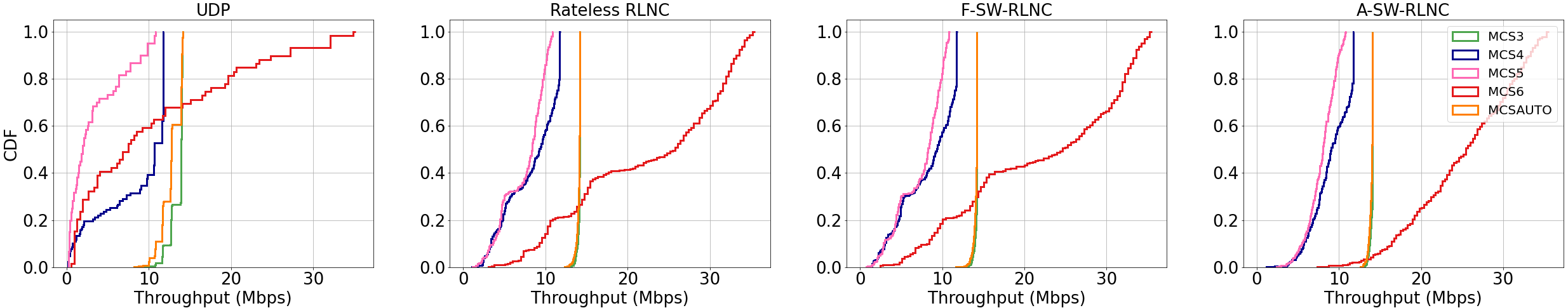}
	\caption{Normalized throughput for various communication solutions.}
	\label{fig:through_res}
\end{figure*}

%\subsection{Relation of Modulation Coding Schemes and Packet Loss }\label{subsubsec:loss}

Fig.~\ref{fig:diff_curves} presents the \gls{rtt} and packet loss event, i.e., the \textit{\gls{cp}}, for several MCS modes obtained in the outdoor mmWave testbed. As expected, packet loss is more prominent for higher MCS values, due to the use of the combination of less reliable modulation (QPSK/BPSK) with a higher \gls{fec} rate. Another interesting finding is the presence of bursty errors (more theoretical insights and validations will be presented in Section~\ref{Sec:FieldEval}). Furthermore, the collected data shows that, locking MCS~3 for both directions and setting the antenna modules to select the MCS mode according to the channel state (MCS Auto) yields similar behaviors when partially obstructed by the metallic plate. Moreover, a high \gls{rtt} variance for MCS 4 through MCS 6 was observed, leading to several issues in the upper layers. This \gls{rtt} fluctuation affects the efficiency of RLNC-based schemes, due to their late reaction to successfully-decoded packets.

\subsection{Emulation: Communication over Transport-Layer mmWave Channel}\label{simulation}

To evaluate the performance of the transport-layer communication solutions under the collected channel profiles \textit{CP}s, an emulator was developed on top of Steinwurf’s Kodo FEC components \cite{kodo_python} (using research license). Our implementation of the Rateless RLNC (sec. \ref{subsec:brlnc}), the Sliding Window variant of the Rateless (F-SW-RLNC), and the A-SW-RLNC (sec. \ref{subsec:acrlnc}) are built upon the \textit{Block} and \textit{Slide} RLNC schemes in these libraries, which we extended accordingly.

The baseline scenario is represented as a UDP transmission of a pseudo-randomized binary file, emulated using the recorded \textit{CP} (see Fig. \ref{fig:diff_curves}). The scenario consists of a file divided into 100 datagrams of 1000 bytes each. Each successful delivery under the defined scenario using a communication algorithm is called an \textit{experience}. The completion of an \textit{experience} outputs a triple, consisting of the normalized throughput, the mean and the maximum in-order delay metrics. A \textit{datapoint} is then defined as the mean of each metric over 10 experiences. The simulation will complete when all time slots from a \textit{CP} are used (\emph{i.e.}, the length of the \textit{CP} sequence is exhausted). The total set of datapoints is then collected per tested \textit{CP} and communication solution. %With respect to the RLNC encoder and the decoder, the generation/maximum window size is set to 5 packets.

\subsection{Performance Analysis of RLNC Solution over mmWave}\label{subsec:rlnc}

After running the emulator with the collected channel profiles, we were able to obtain the main performance metrics for several transport-layer communication solutions. Our results show a significant improvement regarding the in-order delivery delay and normalized data throughput, when network coding is used to improve the communication in the transport layer. Next, we evaluate these performance metrics in detail.

\subsubsection{Throughput}\label{subsubsec:through}

\ifshort

\else
\begin{table*}[h]
\centering
\caption{Statistics for simulation results of tested algorithms from MCS 3 to MCS 6, and Auto modes. For a time slot of $450 \mu$s, the schemes that achieves {\color{DarkGreen} LLC} and {\color{blue} \gls{urllc}} are marked.}
\label{tab:results_mcs6auto}
\begin{tabular}{cccccccccccccc}
\toprule
 &  & \multicolumn{3}{c}{\textbf{Throughput (Mbps)}} & \multicolumn{3}{c}{\textbf{Mean In-Order Delay (slots)}} & \multicolumn{3}{c}{\textbf{Max In-Order Delay (slots)}} \\
\cmidrule(lr){3-3}\cmidrule(lr){4-4}\cmidrule(lr){5-5}\cmidrule(lr){6-6}\cmidrule(lr){7-7}\cmidrule(lr){8-8}\cmidrule(lr){9-9}\cmidrule(lr){10-10}\cmidrule(lr){11-11}\cmidrule(lr){12-12}\cmidrule(lr){13-13}\cmidrule(lr){14-14}
\textbf{Mode} & \textbf{Algorithm} & \textbf{Mean} & \textbf{Stdev}  & \textbf{$P_{99\%}$} & \textbf{Mean} & \textbf{Stdev} & $P_{99\%}$ & \textbf{Mean} & \textbf{Stdev} & \textbf{$P_{99\%}$} \\
\toprule
\multirow{4}{*}{MCS3} & UDP transmission & 13.57 & 0.86 & 14.17 & 6.29 & 7.74 & 32.95 & 8.95 & 7.74 & 35.67 \\
 & Rateless RLNC & 14.08 & 0.35 & 14.22 & 37.67 & 2.64 & 51.62 & 42.52 & 2.84 & 40.80 \\
%& Rateless (new) & 14.07 & 0.30 & 14.22 & 21.16 & 0.42 & 22.24 & 23.07 & 1.17 & 27.32 \\
 &  {\color{DarkGreen} F-SW-RLNC} & 14.15 & 0.18 & 14.22 & {\color{DarkGreen} 3.85 } & 1.97 & {12.11} & {8.13} & 4.24 & { 24.21} \\
 & {\color{blue} A-SW-RLNC } & 14.00 & 0.19 & 14.08 & { 3.11 }& 0.29 & {4.31} & 3.86 & 1.95 & {\color{blue} 11.38} \\
 %& {\color{blue} A-SW-RLNC FF95\% } & 14.00 & 0.51 & 14.08 & { 3.04} & 0.15 & {3.61} & 3.72 & 1.71 & {\color{blue} 10.83} \\
 %& {\color{blue}  A-SW-RLNC FF99\% } & 14.00 & 0.18 & 14.08 & { 3.09} & 0.23 & {4.08} & 3.91 & 1.96 & {\color{blue} 11.78} \\
\hline
\multirow{4}{*}{MCS4} & UDP transmission & 8.62 & 4.27 & 0.23 & 425.44 & 1180.34 & 5106.57 & 427.41 & 1180.38 & 5108.70 \\
 & Rateless RLNC & 8.10 & 3.59 & 1.40 & 140.53 & 180.12 & 765.62 & 148.18 & 181.56 & 779.55 \\
 %& Rateless (new) & 8.41 & 3.40 & 1.87 & 67.98 & 87.83 & 346.58 & 82.69 & 97.02 & 377.86 \\
 &  F-SW-RLNC & 8.33 & 3.39 & 1.59 & 114.29 & 163.71 & 681.71 & 145.74 & 175.30 & 734.11 \\
 &  {\color{DarkGreen}A-SW-RLNC }& 9.11 & 2.36 & 3.87 & {\color{DarkGreen} 16.66} & 20.04 & 80.23 & {63.86} & 77.37 & 324.84 \\
% & {\color{DarkGreen}  A-SW-RLNC FF95\% }& 9.20 & 2.23 & 3.94 & {\color{DarkGreen} 9.85} & 10.81 & 47.56 & {53.15} & 64.10 & 266.22 \\
% & {\color{DarkGreen}  A-SW-RLNC FF99\% } & 9.16 & 2.23 & 4.26 & {\color{DarkGreen} 12.38 } & 13.71 & 61.89 & {54.54} & 62.97 & 271.85 \\
\hline
\multirow{4}{*}{MCS5} & UDP transmission & 3.37 & 3.32 & 0.22 & 1004.99 & 1300.52 & 4980.49 & 1007.07 & 1300.57 & 4982.72 \\
 & Rateless RLNC & 7.07 & 2.87 & 1.23 & 137.53 & 178.17 & 805.66 & 146.27 & 179.22 & 820.97 \\
 %& Rateless (new) & 7.42 & 2.60 & 1.78 & 69.19 & 85.92 & 414.00 & 82.97 & 91.68 & 451.76 \\
 &  F-SW-RLNC & 7.40 & 2.74 & 1.57 & 115.35 & 149.07 & 635.49 & 156.07 & 154.91 & 688.98 \\
 &  {\color{DarkGreen} A-SW-RLNC}  & 7.93 & 1.72 & 3.68 & {\color{DarkGreen} 17.72} & 19.50 & 98.44 & {68.36} & 73.52 & 328.87 \\
% & {\color{DarkGreen}  A-SW-RLNC FF95\% }& 8.10 & 1.68 & 3.54 & {\color{DarkGreen} 10.04} & 10.29 & 47.87 & {56.88} & 64.30 & 294.74 \\
% &  {\color{DarkGreen} A-SW-RLNC FF99\% }& 7.97 & 1.66 & 3.72 & {\color{DarkGreen} 13.16 }& 12.37 & 59.35 & {59.78} & 57.82 & 281.64 \\
\hline
\multirow{4}{*}{MCS6} & UDP transmission & 10.64 & 9.78 & 0.80 & 1029.80 & 1299.62 & 4601.51 & 1031.76 & 1299.59 & 4604.08 \\
 & Rateless RLNC& 21.39 & 10.36 & 3.45 & 177.93 & 228.41 & 999.80 & 186.20 & 229.38 & 1009.81 \\
 %& Rateless (new) & 20.51 & 9.63 & 3.20 & 91.25 & 117.02 & 486.22 & 108.12 & 124.39 & 518.00 \\
 &  F-SW-RLNC & 22.04 & 10.07 & 3.97 & 154.31 & 192.85 & 848.23 & 193.42 & 199.65 & 899.50 \\
 &   { \color{DarkGreen}A-SW-RLNC} & 25.17 & 6.36 & 11.01 & {\color{DarkGreen}  22.28} & 23.63 & 100.38 & {87.16} & 88.85 & 405.18 \\
% & {\color{DarkGreen} A-SW-RLNC FF95\% }& 25.70 & 6.12 & 10.81 & {\color{DarkGreen} 11.77 }& 11.47 & 49.37 & {70.03} & 71.46 & 294.64 \\
% &  {\color{DarkGreen} A-SW-RLNC FF99\% }& 25.57 & 5.95 & 12.13 & {\color{DarkGreen} 15.72 }& 14.53 & 59.23 & {72.28} & 66.27 & 267.64 \\
\hline
\multirow{4}{*}{Auto} & UDP transmission & 12.82 & 1.21 & 14.15 & 12.96 & 11.63 & 43.20 & 15.60 & 11.63 & 45.82 \\
 & Rateless RLNC& 13.98 & 0.50 & 14.22 & 38.31 & 4.03 & 55.08 & 43.27 & 4.20 & 61.43 \\
% & Rateless (new) & 13.81 & 0.61 & 14.22 & 21.24 & 0.60 & 22.95 & 23.44 & 1.73 & 30.59 \\
 & {\color{DarkGreen} F-SW-RLNC} & 14.08 & 0.27 & 14.22 & {\color{DarkGreen} 5.15} & 3.29 &  15.43 & {11.25} & 6.48 & 29.90 \\
 &  {\color{blue} A-SW-RLNC} & 13.92 & 0.24 & 14.08 & { 3.22} & 0.42 & {4.90} & 4.78 & 2.71 & {\color{blue} 14.85} \\
 %&  {\color{blue} A-SW-RLNC FF95\% }& 13.93 & 0.53 & 14.08 & {3.08} & 0.18 & {3.76} & 4.46 & 2.49 & {\color{blue} 14.80 }\\
 %&  {\color{blue} A-SW-RLNC FF99\% }& 13.90 & 0.54 & 14.08 & { 3.19} & 0.34 & {4.57} & 4.97 & 2.78 & {\color{blue} 14.25} \\
\bottomrule
\end{tabular}
\end{table*}
\fi

% \begin{table}[h]
%     \centering
%     \caption{1st percentile throughput values for simulation of RLNC algorithms.}
%     \label{tab:through_1pp}
%     \begin{tabular}{lcccccccc}
%         \multicolumn{1}{c}{}                                        & \multicolumn{2}{c}{\textbf{Rateless}} & \multicolumn{2}{c}{\textbf{ F-SW-RLNC}} & \multicolumn{2}{c}{\textbf{ A-SW-RLNC}} \\
%         \multicolumn{1}{c}{\multirow{-2}{*}{\textbf{$P_{1\%}$}}}   & \textbf{Auto}            & \textbf{MCS6}           & \textbf{Auto}            & \textbf{MCS6}           & \textbf{Auto}     & \textbf{MCS6}    \\
%         \hline
%         $\eta$ (Mbps)   & 11.86                    & 3.44                   & 13.03                    & 3.97                   & 12.98             & 11.00
%        \end{tabular}
% \end{table}

The results in Fig.~\ref{fig:through_res} show that the robustness of the A-SW-RLNC over the mmWave network link pertains a higher overall throughput compared to the UDP baseline across all MCS modes, despite the high RTT variance. Besides, higher MCS modes obtain higher throughput gains as expected.
%, as per time-slot feedback permits state synchronization between the encoder and the decoder for each transmitted packet, culminating in 100\% delivery success for each experience.
For MCS 3 and MCS Auto, A-SW-RLNC tops other coding solutions only slightly ahead, and at the same time presents a consistent behaviour (\emph{i.e.}, less fluctuations). Table~\ref{tab:results_mcs6auto} shows that, for the MCS Auto mode, there is no notable difference on the 99$^{th}$ percentile between the Rateless-RLNC and the A-SW-RLNC. This means that A-SW-RLNC shows no significant throughput penalty while, as we show later, it results in significant delay improvements.

% \subsection{Experimental mmWave Network Test Scenario }\label{section:testbed}
% \begin{figure*}[!h]
%  	\centering
% 	\caption{Maximum in-order delay results.}
% 	\label{fig:maxdelay_res}
% \end{figure*}

\begin{figure*}[h]
	\centering
	\includegraphics[width=\textwidth,keepaspectratio]{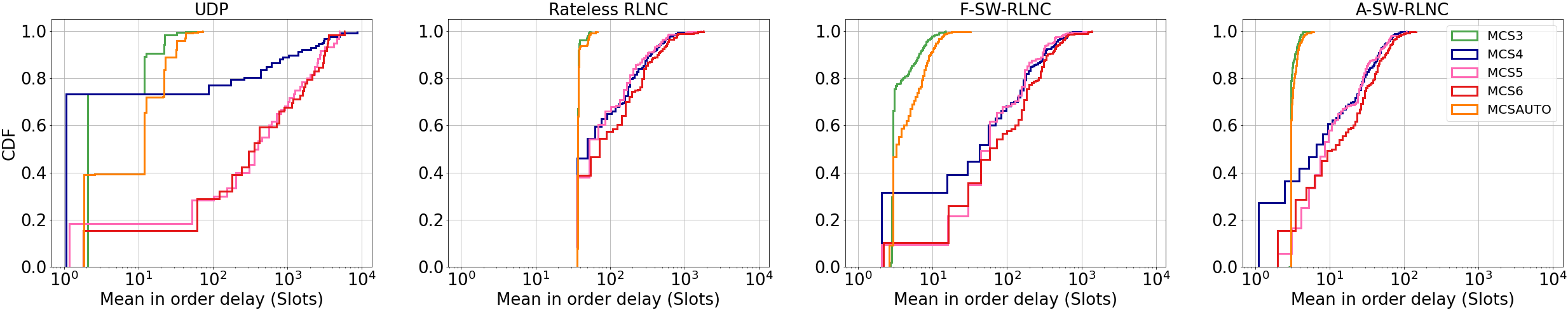}
	\ifshort\else
	\\\vspace{0.2cm}
	\includegraphics[width=\textwidth,keepaspectratio]{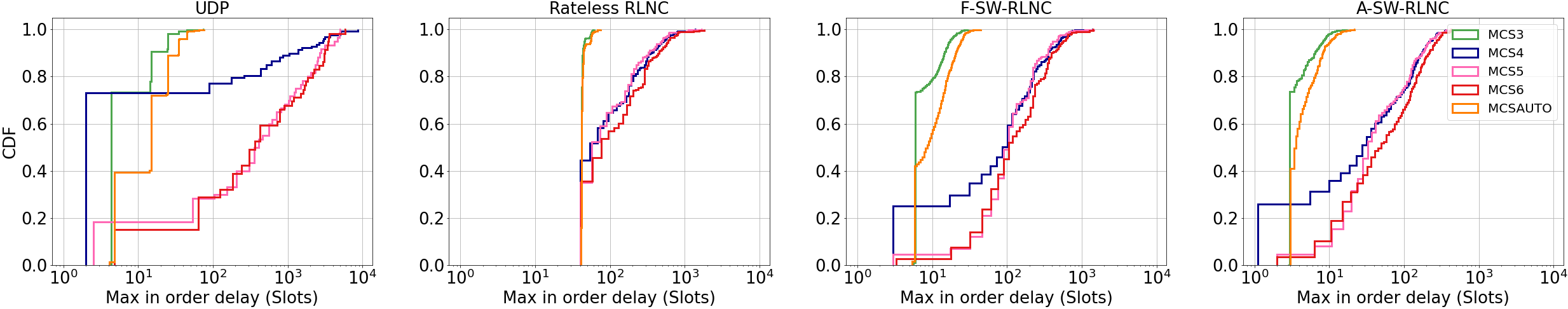}
	\fi

	\caption{Top and bottom rows present the empirical distribution for the mean in-order delay and max in-order delay, respectively, obtained via the collected experiences for each communication solution (UDP, R-RLNC, F-SW-RLNC and A-SW-RLNC) and each MCS mode.}
	\label{fig:delay_res}
\end{figure*}

\subsubsection{Mean In-Order Delivery Delay}\label{subsubsec:mean_delay}

% \begin{table}[h]
%     \centering
%     \caption{99th percentile values for simulation of RLNC algorithms.}
%     \label{tab:rlnc_99pp}
%     \begin{tabular}{lcccccccc}
%         \multicolumn{1}{c}{}                                        & \multicolumn{2}{c}{\textbf{Rateless}} & \multicolumn{2}{c}{\textbf{ F-SW-RLNC}} & \multicolumn{2}{c}{\textbf{ A-SW-RLNC}} \\
%         \multicolumn{1}{c}{\multirow{-2}{*}{\textbf{$P_{99\%}$}}}   & \textbf{Auto}            & \textbf{MCS6}           & \textbf{Auto}            & \textbf{MCS6}           & \textbf{Auto}     & \textbf{MCS6}    \\
%         \hline
%         $\eta$ (Mbps)   & 14.22                    & 35.13                   & 14.22                    & 35.52                   & 14.08             & 35.55            \\
%         $D_{mean}$ (t)  & 55.08                    & 999.80                  & 15.42                    & 848.22                  & 4.90              & 100.38           \\
%         $D_{max}$ (t) & 61.43                    & 1009.81                 & 29.9                     & 899.50                  & 14.85             & 405.18
%         \end{tabular}
% \end{table}

Our results show that the A-SW-RLNC solution performs exceptionally well regarding in-order delivery delays in comparison with the simpler rateless solutions over the mmWave channel profiles, as illustrated in Fig.~\ref{fig:delay_res}.
Regarding the mean in-order delay, with MCS 3 and MCS Auto modes, CDF curves show that the simple UDP transmission achieves lower delays than the R-RLNC scheme. In R-RLNC, a batch of packets is decodable only if the decoder receives a number of coded packets equal to the number of original packets in the batch (Section~\ref{subsec:brlnc}). The sender, thus, keeps sending batches of encoded data for the same generation until it ensures the batch can be decoded, and this limits the minimum theoretical achievable delay.  The F-SW-RLNC implementation removes this limitation and achieves better results for the upper quartile and above with an order of two, see Fig~\ref{fig:delay_res}.

For high MCS modes, the sharp increase in the channel erasure rate makes a simple UDP file transfer to be unusable, as file retransmission probability is substantial, therefore increasing the packet in-order delivery delay. The R-RLNC implementation mitigates these losses via a generous code redundancy, improving the in-order delay of a UDP transmission. From the CDF curves, gains start for the upper steps of the 20$^{th}$ percentile. For the MCS 6, regarding the 99$^{th}$ percentile of the mean in-order delay, the R-RLNC outperforms the UDP transmission by a factor of 4.60  (i.e., 4601.5/999.80). %, the .
%Similarly to the lower MCS behavior, although not as prominent, the rateless \textit{slide} RLNC implementation gets a slight performance improvement with respect to the $99\%$ guarantees, as shown in Table \ref{tab:results_mcs6auto} - an increase of $17.8\%$.
Further, for the MCS 6, the A-SW-RLNC outperforms the R-RLNC with a factor of 9.96 (i.e., 999.80/100.38) in the 99$^{th}$ percentile of the mean in-order delay, see Table~\ref{tab:results_mcs6auto}. This order of magnitude improvement over R-RLNC is attained thanks to the the adaptive and dynamic component of A-SW-RLNC (Section~\ref{subsec:acrlnc}).
%A-SW-RLNC gets highly ahead in this regard: the adaptive and dynamic component, as previously mentioned, is now evident from the 99$^{th}$ percentile, outperforming the former algorithm with a factor of $9.96\ (999.80/100.38)$. The obtained guarantee improvement is approximately $10$ times over the R-RLNC coding solution.

\subsubsection{Maximum In-Order Delivery Delay}\label{subsubsec:max_delay}

With respect to the maximum in-order delivery delay, it is shown in Fig.~\ref{fig:delay_res} that the R-RLNC algorithm does not differ much from the previous average delay analysis, showing that its bounds are quite close. %Although penalized by its batch size at low MCS modes (in MCS3 and MCS Auto), as shown in Table~\ref{tab:results_mcs6auto},
For the MCS 6 and in the 99$^{th}$ percentile, the statistical values for the maximum delay show a significant improvement with a factor of 4.22 (i.e., 4604.08/1009.81) times for R-RLNC compared to UDP transmission and 11.36 (i.e., 4604.08/405.18) times for A-SW-RLNC compared to UDP transmission, respectively.
%and 5.54 (i.e.. 1031.76/186.20) times for the mean values at MCS6.
Similar to the average analysis, the F-SW-RLNC shows a slight improvement in the maximum in-order delivery delay. A-SW-RLNC achieves an improvement with a ratio of 2 to 2.5 across all percentile bounds over the former approach (F-SW-RLNC), from MCS Auto to MCS 6. %Moreover, compared to the baseline UDP transmission, at MCS6 and for the 99\% maximum delay,
%the adaptive and causal algorithm results in much lower guarantees, with a decrease over 11.36 (i.e., 4604.08/405.18) times.

\subsubsection{{\color{DarkGreen} LLC} and {\color{blue} URLLC} Performance Indicators}\label{subsubsec:llc_urlc}
Regarding the support of low latency and ultra-reliable scenarios in the mmWave IAB, Table \ref{tab:results_mcs6auto} highlights the schemes that are capable to achieve {\color{DarkGreen} LLC} and {\color{blue} URLLC} requirements. For {\color{DarkGreen} LLC} applications, we target a mean in-order delay below 10~ms (22 slots).
%, and a max in-order delay ($P_{99\%}$) below 30ms (67 slots).
For {\color{blue} URLLC} applications (only $P_{99\%}$),
%a mean in-order delay below 5ms (11 slots) and
a max in-order delay below 10~ms (22 slots) is targeted. As presented in Table II, only the A-SW-RLNC can support {\color{blue} URLLC} applications, at MCS 3 and MCS Auto modes, by obtaining
%a mean in-order delay for $P_{99\%}$ below 5ms (auto and MCS3), and at the same time,
a  max in-order delay below 10~ms for $P_{99\%}$. In such lossy links, transport protocols like UDP cannot be used in {\color{blue} URLLC} scenarios, as well as the rateless and the F-SW-RLNC schemes. When addressing {\color{DarkGreen} LLC} applications,
%the A-SW-RLNC scheme presents a huge improvement, compared to all the other schemes, and also with the dynamic approach used by lower layers (auto).
the A-SW-RLNC scheme is capable to achieve a delay below 10~ms in all the MCSs evaluated, allowing the increase of the network bandwidth by using a higher MCS. As presented, low-layer techniques are very conservative and do not allow to take the full benefits of the mmWave link capacity.

%\begin{figure}[h]
%\includegraphics[width=0.5\textwidth]{fig/RLNC UDP Results/First results (MCS)/graph_script_and_results_Graphs_success_rate.png}
%\caption{MCS Modes vs RLNC}
%\label{fig:}
%\end{figure}

%\begin{figure}[h]
%\includegraphics[width=0.5\textwidth]{fig/RLNC UDP Results/Second Results (MCS Auto)/process_time_eq100.png}
%\caption{Processing time}
%\label{fig:}
%\end{figure}

%\begin{figure}[h]
%\includegraphics[width=0.5\textwidth]{fig/RLNC UDP Results/Second Results (MCS Auto)/packet_overhead_eq100.png}
%\caption{Packet Overhead}
%\label{fig:}
%\end{figure}

%\begin{figure}[h]
%\includegraphics[width=0.5\textwidth]{fig/RLNC UDP Results/Second Results (MCS Auto)/success_rate.png}
%\caption{Success Rate}
%\label{fig:}
%\end{figure}

%%%%%%%%%%%%%%%%%%%%%%%%%%%%%%%%%%%%%%%%%%%%%%%%%%%%%
\section{Theoretical Validations}\label{sec:theoretical}
%%%%%%%%%%%%%%%%%%%%%%%%%%%%%%%%%%%%%%%%%%%%%%%%%%%%%

In this section, we perform theoretical validations for the field experiments of Section~\ref{sec:experiments}. In the first subsection, we investigate the packet loss sequence for the transport-layer mmWave channel, and demonstrate that it can be well-modeled with a Gilbert-Elliott (GE) channel with erasures \cite{gilbert1960capacity,sadeghi2008finite}. Obtaining the mathematical model for the channel enabled us to derive upper-bounds on the in-order delivery delay of our communication solution, in the second subsection, which is very helpful to define delay guarantees.

\subsection{Modeling the Transport mmWave Channel}\label{Sec:FieldEval}

\begin{figure}
    \centering    \includegraphics[width=1 \columnwidth]{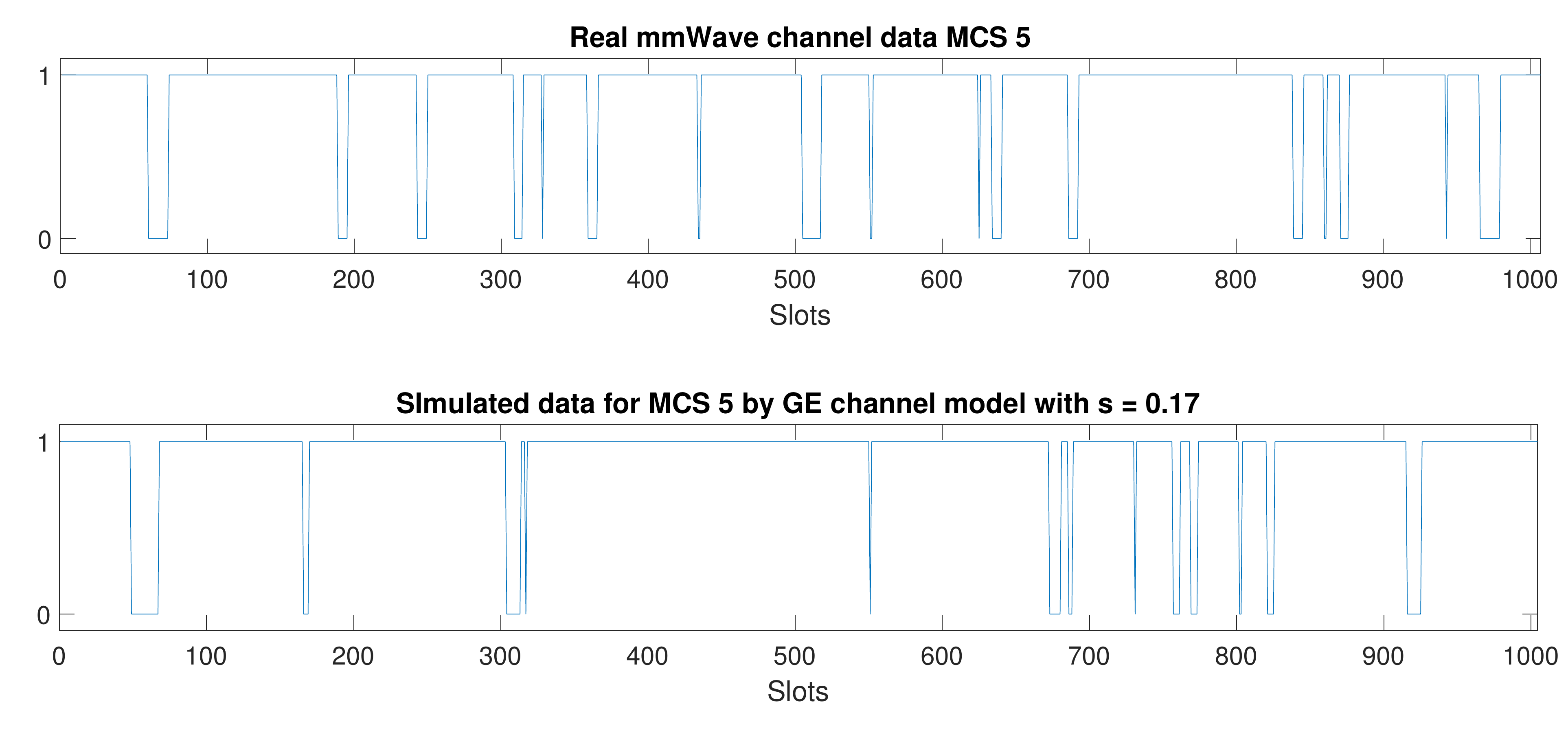}
    \caption{The sequence of packet loss realizations for MCS mode $5$: (top) recorded data from the testbed, (bottom) recorded data from the GE simulator and $s=0.17$.}
    \label{fig:GEvsMCS}
\end{figure}

Here, we take a closer look at the nature of errors observed in the testbed experiments in Section~\ref{sec:experiments}. The point-to-point packet loss pattern observed in the transport layer mmWave channel has a bursty behaviour, resulting in a contiguous sequence of packets to be erased in the case of failure. The top panel of Fig.~\ref{fig:GEvsMCS} shows a binary sequence that indicates whether or not a packet is lost at a time slot, for the MCS~5, according to the field data. The same bursty behaviour was observed across all the MCS modes. Such behavior was already shown in \cite{Dahhani2019} and can be explained by the LOS characteristics of the mmWave. When the link is blocked, the connection between the two stations is lost, and the Sector Level Sweep (SLS) mechanism searches for a new pair of sectors capable to restore the link. The search time depends not only on the interference nature, but also on the size of the array of antennas, which could exceed the timeout.% to send the packet.

Therefore, we model the point-to-point mmWave channel, observed from the transport layer, as a Gilbert-Elliott (GE) channel with erasures. The channel model is a Markov process with two states, i.e., good and bad. In the good state, the erasure probability is zero, and in the bad state, the erasure probability is one. We denote with $s$ and $q$, the transition probabilities from good state to bad state, and from the bad state to good state, respectively. The stationary distribution of the GE channel is given by $\pi_G = {s}/(s+q)$ and $\pi_B = 1 - \pi_G$.
%In the simulated channel beaver, $\epsilon_G = 0$ and $\epsilon_B = 1$ denote the average erasure rate at the corresponding states. Such that,
The average erasure rate is $\epsilon_\text{mean} = \pi_B$, and  $1/s$ is the average erasure burst. By computing the average erasure rate and the average erasure burst, from the data we collected in our experiments for MCS~5, we set $s=0.17$ and $\epsilon = 0.1$. Hence, the transition probability from the bad state to good state is given by $q={s\pi_B}/{\pi_G}=0.019$.
The bottom panel of Fig.~\ref{fig:GEvsMCS} illustrates the packet loss sequence, for the MCS~5 mode, according to the GE channel model.
%We collected the information for $4942$ data packets delivered in-order form $6863$ transmissions over the mmWave channel, i.e., average rate of $0.72$. Then,

\begin{figure}
    \centering
    \includegraphics[width=0.99 \columnwidth]{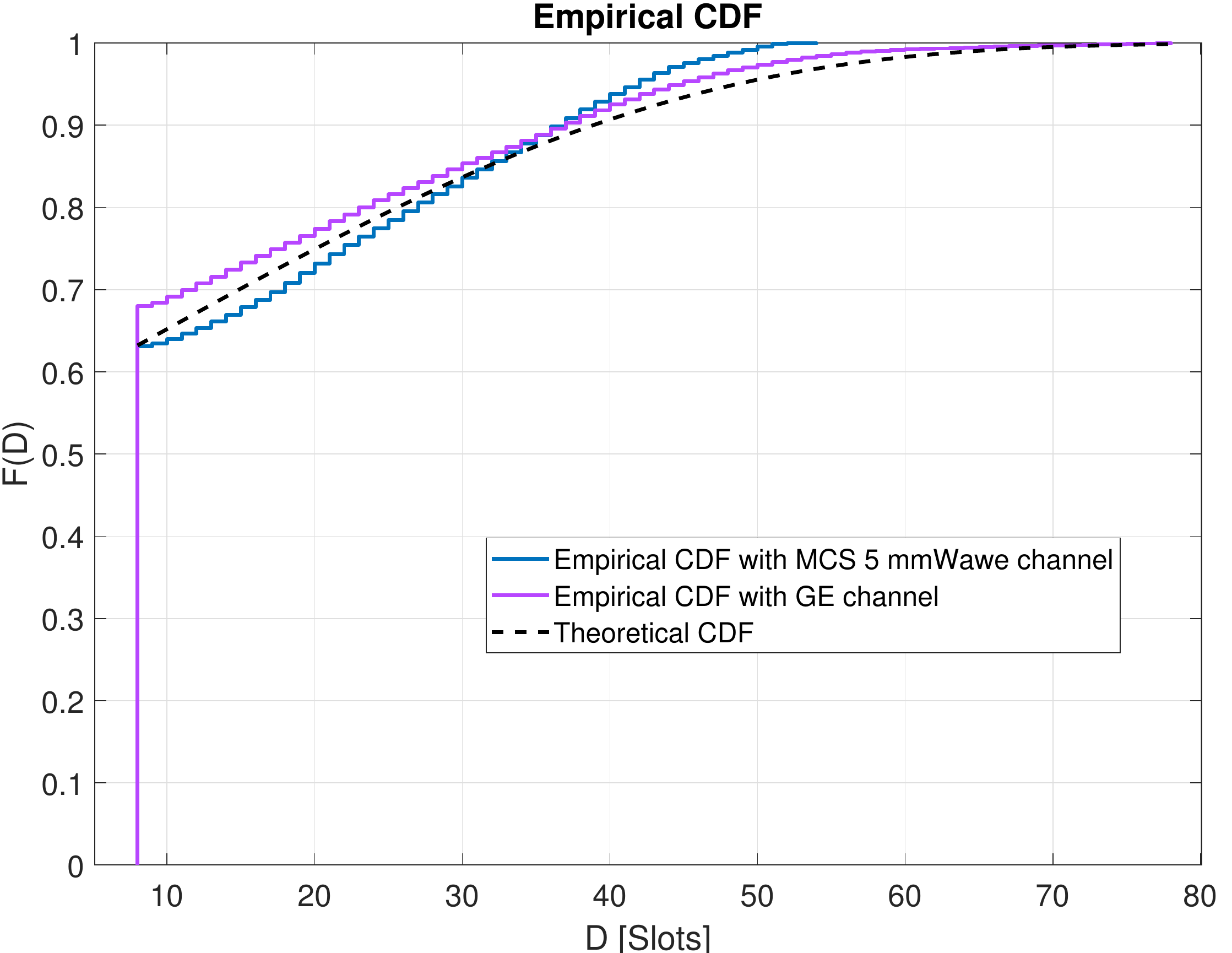}
    \caption{Empirical CDF of the in-order delivery delay.}
    \label{fig:simvsexpcdf}
\end{figure}

We then use the A-SW-RLNC solution \cite{cohen2020adaptive} simulated over GE channel with the aforementioned parameters and $\text{RTT}=16$. Fig.~\ref{fig:simvsexpcdf} shows the empirical CDF of the in-order delivery delay for the two sets of data. As we observe, there is a good agreement between the empirical distribution of delays, based on the simulator and the real-field mmWave data, which validates that the GE channel is the appropriate model to approximate the bursty mmWave channel we observed from the transport layer. The theoretical curve with dashed line shows the CDF for the type~1 extreme value distribution with location and scale parameters equivalent to the empirical mean and standard deviation of delays. It is used to model the distribution of the maximum of a number of samples.

%\begin{figure}
%    \centering
%    \includegraphics[width=0.99 \columnwidth]{CDF_v3.eps}
%    \caption{Empirical CDF of the in-order delivery delay.}
%    \label{fig:simvsexpcdf}
%\end{figure}

\subsection{Delay Analysis of A-SW-RLNC Solution with Cross-Layer Considerations}

In this subsection, we investigate the statistical properties of the in-order delay for the transport-layer A-SW-RLNC solution over the GE channel model, to provide theoretical delay guarantees. Here, we follow similar techniques as used in \cite{cohen2020adaptive,cohen2022broadcast}, with the difference that instead of considering the average failure rate, we consider an extreme value, such that most of the time the erasure rate falls below this value. Such extreme analysis is necessary to provide delay guarantees required by the \gls{urllc}.

The transport-layer estimation of the average erasure probability at the $t$-th time slot, as represented by the feedback acknowledgements from the MAC layer \cite{Dahhani2019,chen2012throughput,li2005performance,hiertz2005throughput}, is given by
\begin{equation*}
    \epsilon_{\text{mean}} = 1-\frac{\sum_{j=1}^{t-\text{RTT}} u_j}{t-\text{RTT}}.
\end{equation*}
Here, $u_j$ is the binary feedback acknowledgement at the time slot $j$. We note that, for tracking the channel rate, only the number of positive ACKs matters, and their association with particular packet indices (being in-order or not) is not needed.

The erasure rate is time-variant for the GE channel, which affects the bounds we can define on the in-order delivery delay. Practically, using its average value and the standard deviation, we can obtain an upper-bound on the erasure rate, such that, most of the time the erasure probability falls below this bound, i.e.,
\begin{equation}\label{eq:e_max}
    \epsilon_{\text{max}}^{\alpha} = \epsilon_{\text{mean}} + \alpha \frac{\sqrt{\nu}}{\text{RTT}},
\end{equation}
where $\nu$ is the variance of the channel during one RTT period, and $\alpha$ is the confidence factor, using the deviation rule. Therefore, we use this conservative measure of the erasure probability to obtain an upper bound on the delay with some strong guarantees to hold.

As elaborated in Section~\ref{Sec:FieldEval}, the behavior of the bursty transport layer channel can be represented by using a GE channel model with erasures. For $\epsilon_G$ (erasure probability in the good state) and $\epsilon_B$ (erasure probability in the bad state) the average erasure rate is given by $\epsilon_\text{mean}=\pi_B\epsilon_B=\epsilon_B{q}/({q+s})$, and the variance by
\begin{equation*}
  \nu = \left((\pi_B \epsilon_B)-(\pi_B \epsilon_B)^2\right)\text{RTT}.
\end{equation*}
We note that the delay guarantees provided can be considered for other channel models using their mean and variance statistics, in a similar fashion.

By incorporating the extreme-case erasure rate  $\epsilon_{\text{max}}^{\alpha}$ in the retransmission condition of the A-SW-RLNC solution, we obtain
\begin{equation}\label{eq:part1}
1 - d  -\epsilon_{\text{max}}^{\alpha} > th.
\end{equation}
Here, $d$ is the rate of degrees of freedom (Dofs), i.e., the ratio of the number of DoFs needed to decode the transmitted coded packets (see \eqref{eq:ac_coded}) and the number
of DoFs added by apriori and aposteriori FECs, and $th$ is a threshold influencing the throughput-delay trade-off. The retransmission criteria appears in probabilistic expressions, e.g., retransmission probability, that determine an upper bound on the mean in-order delay in \cite[Section V.A]{cohen2020adaptive}. Thus, by considering the extreme-case erasure rate $\epsilon_{\text{max}}^{\alpha}$, we obtain bounds on these probabilities.

For the upper bound on the in-order delivery delay in \cite{cohen2020adaptive}, it also appears a scaling term which indicates the maximum number of retransmissions needed to succeed in the forward channel. For the GE channel model, this upper bound on the scaling factor is given by
\begin{equation}\label{eq:part2}
    \left(\frac{1}{1-s}\right)\left(\frac{1}{s}-s\right)-1.
\end{equation}
Using the two considerations (\ref{eq:part1}) and (\ref{eq:part2}), and the similar techniques as in \cite{cohen2020adaptive}, we derive upper-bound $D_\text{up}^\alpha$ on the in-order delivery delay, that holds with a high probability (rather than the mean in-order delay). This probability is tune-able using the confidence parameter $\alpha$ in (\ref{eq:e_max}).

%for the maximum number of retransmissions needed to succeed in the GE channel under the selection of $\alpha$, we obtain directly an in-order delay upper bound using equations (6), (7) and (8) in \cite[Section V.A]{cohen2020adaptive}.
%Thus, selecting the standard deviation factor parameter, $\alpha$ and using the cumulative distribution function the mean in-order delivery delay is given by
%\begin{multline*}
%D_{up}^{\alpha}(\epsilon_{\text{max}}^{\alpha}) \leq \lambda {D_{up}^{\alpha}}(\epsilon_{\text{max}}^{\alpha})_{[no\mbox{ }feedback]}\\
%+(1-\lambda )\Big({D_{up}^{\alpha}}(\epsilon_{\text{max}}^{\alpha})_{[nack\mbox{ }feedback]}\\
%+{D_{up}^{\alpha}}(\epsilon_{\text{max}}^{\alpha})_{[ack\mbox{ }feedback]}\Big),
%\end{multline*}
%where ${D_{up}^{\alpha}}_{[\cdot]}$ denote the delay upper bounds for GE channel with selected $\alpha$ under different feedback states and $\lambda$ denote the normalization parameter for the fraction of time without feedback.

\begin{figure}
    \centering
    \includegraphics[width=0.99 \columnwidth]{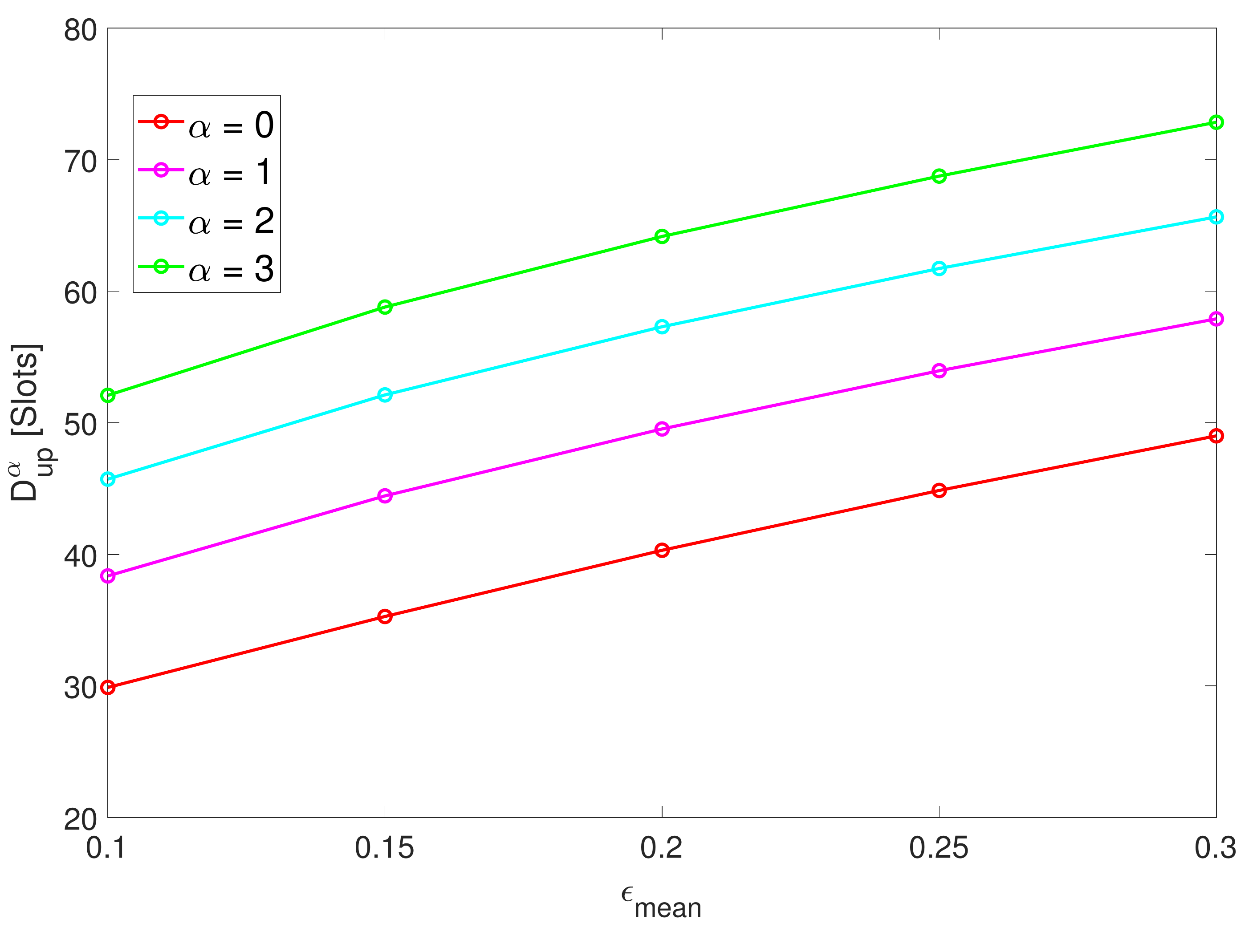}
    \caption{Upper-bounds on the in-order delivery delay as a function of the average erasure rate $\epsilon_\text{mean}=\pi_B$ for several values of the confidence parameter $\alpha$.}
    \label{fig:d_up}
\end{figure}

Next, we evaluate the derived bounding methodology on our collected field data and simulated data. Fig.~\ref{fig:d_up} shows the upper bound on the in-order delivery delay for several values of $\alpha$ as a function of the average erasure rate $\epsilon_\text{mean}=\pi_B$ and for $s=0.17$, given $\epsilon_\text{mean}=\pi_B$, $q=s\pi_B/(1-\pi_B)$. As we observe, the probabilistic upper bound increases, by increasing $\alpha$, however, with a higher probability to be satisfied.

The theoretical analysis performed in this section helped us define bounds on the delay, such that the observed delays fall below these bounds with a high probability. This probability, which we call probability of success $P_{suc}^{\alpha}$, depends on the $\alpha$, the confidence factor we considered when we deviated from the average value of erasure rate in our analysis. By increasing $\alpha$, the upper-bound also increases, but it holds with a higher probability. In fact, each value of $\alpha$ corresponds to an upper-bound of the in-order delivery delay (horizontal axis of Fig~\ref{fig:simvsexpcdf}), and $P_{suc}^{\alpha}$ is the value of the CDF function at this delay (vertical axis of Fig~\ref{fig:simvsexpcdf}). As presented in Fig~\ref{fig:d_up}, the theoretical results show that the in-order delay for $P_{99\%}$ ($\alpha=3$), with an $\epsilon_\text{mean}=0.3$ is around 73 slots. This value is close to the empirical results of Section~\ref{subsec:rlnc}, where the MCS 5 mode achieves 98.44 slots for the $P_{99\%}$ mean in-order delay.

\begin{figure*}
    \centering
    \includegraphics[width=1 \textwidth]{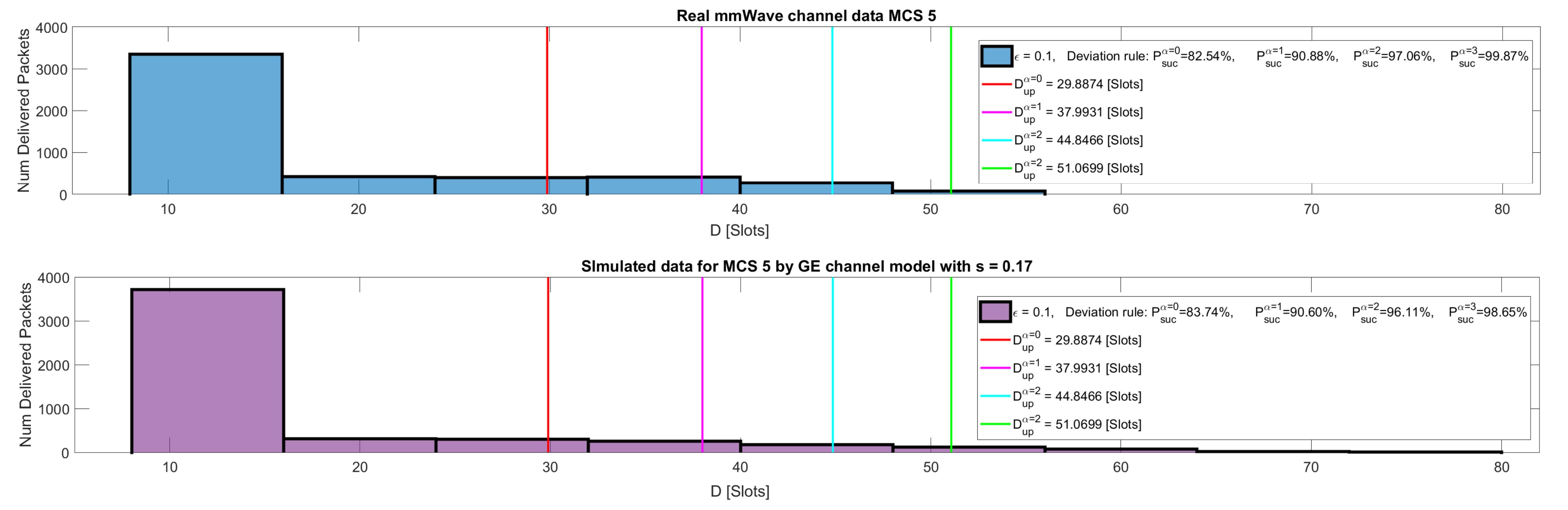}
    \caption{The empirical distribution of the in-order delivery delay obtained via A-SW-RLNC solution and MCS~5 over two channel models, i.e., mmWave testbed (top) and GE simulator (bottom).}
    \label{fig:simvsexpdistvert}
\end{figure*}

Fig.~\ref{fig:simvsexpdistvert} shows the empirical distribution for the in-order delivery delay of the simulated and field data. The derived upper-bound on the delay for several values of $\alpha$ are indicated as vertical lines. The ratio of the points where the in-order delivery delay falls below these vertical lines, i.e., $P_{suc}^{\alpha}$, appear in the legend. We highlight that changing the view from the average analysis of delay to the worst-case scenario is particularly important for \gls{urllc} applications, where guarantees are demanded on the in-order delivery delay.

%The support of LLC services in MCS5 could be achieved in the emulation scenario $-17.72$ slots ($10ms=22\ slots$). Comparing with the theoretical validation, for an $\epsilon_\text{mean}=0.1$, the upper-bound could be 30 slots, which could be explained by the fixed erasure rate used.

%%%%%%%%%%%%%%%%%%%%%%%%%%%%%%%%%%%%%%%%%%%%%%%%%%%%%
\section{Conclusions and Future Visions}\label{sec:conclusions}
%%%%%%%%%%%%%%%%%%%%%%%%%%%%%%%%%%%%%%%%%%%%%%%%%%%%%
In this work, we proposed a significant enhancement on the mmWave performance by incorporating network coding algorithms to stabilize the high-frequency communication sensitivity. %This sensitivity caused due to high co-channel interference, lack of line of sight, and low propagation and penetration of the signal in outdoor environments result in high losses and high latency.
%While classical solutions today to obtain reliable communication pay in high redundancy in the PHY layer, which reduces the effective bandwidth dramatically, network coding by tracking the current pattern of channel condition can compensate the losses efficiently\off{ without data-rates suffering}.
In particular, we showed that, using A-SW-RLNC, it is possible to obtain ultra-reliable high bandwidth while reducing by up to an order of two the mean in-order delay. %Today’s technological world demands effective ultra-reliable low-latency communications to handle large volume of data. Along this trajectory, we envision several line of work that can built upon and benefit from the proposed technology in this paper. First, we would like to highlight that
Our results demonstrate that the communication protocols can notably take benefit from relaxing the PHY and MAC layer error control mechanisms, and delegating the task to the upper layers using the proposed network coding solution. In fact, the retransmissions that occur due to MAC error control mechanisms are effectively not needed once the network coding solution and FEC mechanisms are utilized.%, as our field results also verify.

As for future work, we plan to use the gain of \gls{mp} network coding communication through splitting the mmWave band into several sub-bands. For this end, we will extend the proposed \gls{sp} solution to several frequency links via an effective MP coded communication \cite{cohen2020adaptiveMPMH}. % to offer outstanding throughput-delay performance, demanded by new generation of technologies. The MP solution are targeted to be fully implemented in UDP \cite{} and in QUIC \cite{cohen2021bringing}. As we envision
%Additionally, we also want to evaluate the use of the proposed solution in the MAC layer, to solve the additional delay caused by the current block-ack scheme used by the standard. We believe that introducing such changes in the MAC layer will help to fix some of the issues that current transport protocols have in mmWave networks.
To use the proposed solution over highly-meshed backhaul of novel communication networks, we plan to incorporate software-defined controllers for collecting information that can enhance the communication performance over meshed mmWave communication \cite{bib:Cohen21}. Last but not the least, we plan to exploit the recent trend in the estimation of error patterns using deep-learning solutions to further improve our adaptive solutions over mmWave networks \cite{cohen2021deepnp}.

%Future work includes the:
%\ale{
%\begin{itemize}
    %\item Extension of the proposed single-path (SP) solution as function of the number of frequency links. Incorporating the relaxation in the PHY layer mechanism's.
    %\item Full MP implementation in UDP and possibly in QUIC as presented in \cite{cohen2021bringing}.
    %\item Controller for collecting information that can enhance mmWave performance.
    %\item Prediction of mmWave channel pattern by deep-learning solutions as given in \cite{cohen2021deepnp}.
%\end{itemize}
%}

\section{Acknowledgments}
This work is supported by the European Regional Development Fund (FEDER), through the Regional Operational Programme of Centre (CENTRO 2020) of the Portugal 2020 framework and FCT under the MIT Portugal Program [Project SNOB-5G with Nr. 045929(CENTRO-01-0247-FEDER-045929)].

%%%%%%%%%%%%%%%%%%%%%%%%%%%%%%%%%%%%%%%%%%%%%%%%%%%%%
\bibliographystyle{IEEEtran}
\bibliography{references}
%%%%%%%%%%%%%%%%%%%%%%%%%%%%%%%%%%%%%%%%%%%%%%%%%%%%%
\end{document}